\documentclass[pra,superscriptaddress,twocolumn,bibliography]{revtex4-1}

\usepackage{multirow}
\usepackage{relsize}

\usepackage{enumerate}
\usepackage{amsmath}
\usepackage{amssymb}
\usepackage{graphicx}
\usepackage{bm}
\usepackage{color}
\usepackage{hhline}
\usepackage{theorem}
\usepackage{mathtools}
\usepackage[colorlinks]{hyperref}

\newtheorem{Lemma}{Lemma}

\newcommand{\cA}{{\mathcal{A}}}

\newcommand{\cH}{{\mathcal H}}

\newcommand{\cT}{{\cal T}}

\newcommand{\cE}{\mathcal{E}}

\newcommand{\cN}{{\mathcal N}}

\newcommand{\cD}{{\mathcal D}}

\newcommand{\mc}[1]{\mathcal{#1}}
\newcommand{\ket}[1]{|#1\rangle} 
\newcommand{\bra}[1]{\langle#1|} 
\newcommand{\ketbra}[2]{|#1\rangle\langle#2|} 

\DeclareMathOperator{\tr}{Tr}

{\theorembodyfont{\normalfont\it}
\theoremheaderfont{\normalfont\bf}
\newtheorem{thm}{ Theorem}
\newtheorem{dfn}[thm]{ Definition}
\newtheorem{lmm}[thm]{ Lemma}

\newtheorem{crl}[thm]{ Corollary}
\newtheorem{asm}[thm]{ Assumption}
\newtheorem{prp}[thm]{ Proposition}
\newtheorem{cjt}[thm]{ Conjecture}}

{\theorembodyfont{\normalfont}
\theoremheaderfont{\normalfont\it}
\newtheorem{prf}{ Proof:}}

{\theorembodyfont{\normalfont}
\theoremheaderfont{\normalfont\it}
}

{\theorembodyfont{\normalfont}
\theoremheaderfont{\normalfont\it}
}

{\theorembodyfont{\normalfont}
\theoremheaderfont{\normalfont\it}
}

{\theorembodyfont{\normalfont}
\theoremheaderfont{\normalfont\it}
}

{\theorembodyfont{\normalfont}
\theoremheaderfont{\normalfont\it}
}

{\theorembodyfont{\normalfont}
\theoremheaderfont{\normalfont\it}
\newtheorem{rmk}{ Remark.}}

\newcommand{\proj}[1]{\mbox{$\ket{#1}\!\bra{#1}$}}

\newcommand{\alg}[1]{\begin{align}#1\end{align}}

\newcommand{\nn}{\nonumber}
\newcommand{\ca}[1]{{\mathcal #1}}
\newcommand{\mbb}[1]{{\mathbb #1}}

\newcommand{\bthm}[1]{\begin{thm}\label{thm:#1}}
\newcommand{\ethm}{\end{thm}}

\newcommand{\rThm}[1]{Theorem \ref{thm:#1}}
\newcommand{\blmm}[1]{\begin{lmm}\label{lmm:#1}}
\newcommand{\elmm}{\end{lmm}}

\newcommand{\bdfn}[1]{\begin{dfn}\label{dfn:#1}}
\newcommand{\edfn}{\end{dfn}}

\newcommand{\basm}[1]{\begin{asm}\label{asm:#1}}
\newcommand{\easm}{\end{asm}}
\newcommand{\bprp}[1]{\begin{prp}\label{prp:#1}}
\newcommand{\eprp}{\end{prp}}

\newcommand{\bcrl}[1]{\begin{crl}\label{crl:#1}}
\newcommand{\ecrl}{\end{crl}}

\newcommand{\bcjt}[1]{\begin{cjt}\label{cjt:#1}}
\newcommand{\ecjt}{\end{cjt}}

\newcommand{\bprf}{\begin{prf}}
\newcommand{\eprf}{\end{prf}}
\newcommand{\brmk}{\begin{rmk}}
\newcommand{\ermk}{\end{rmk}}
\newcommand{\laeq}[1]{\label{eq:#1}}
\newcommand{\req}[1]{(\ref{eq:#1})}

\newcommand{\bitem}{\begin{itemize}}
\newcommand{\entem}{\end{itemize}}
\newcommand{\benum}{\begin{enumerate}}
\newcommand{\ennum}{\end{enumerate}}

\newcommand{\otm}{\otimes}

\begin{document}

\title{One-shot quantum error correction of classical and quantum information}

\author{Yoshifumi Nakata}
\email{nakata@qi.t.u-tokyo.ac.jp}
\affiliation{Photon Science Center, Graduate School of Engineering,~The University of Tokyo, Bunkyo-ku,~Tokyo 113--8656, Japan}
\affiliation{JST, PRESTO, 4--1--8 Honcho, Kawaguchi, Saitama, 332--0012, Japan}

\author{Eyuri Wakakuwa}
\email{wakakuwa@quest.is.uec.ac.jp}
\affiliation{Department of Communication Engineering and Informatics, Graduate School of Informatics and Engineering, The University of Electro-Communications, Tokyo 182--8585, Japan}

\author{Hayata Yamasaki}
\email{hayata.yamasaki@gmail.com}
\affiliation{Photon Science Center, Graduate School of Engineering,~The University of Tokyo, Bunkyo-ku,~Tokyo 113--8656, Japan}
\affiliation{Institute for Quantum Optics and Quantum Information --- IQOQI Vienna, Austrian Academy of Sciences, Boltzmanngasse 3, 1090 Vienna, Austria}
\affiliation{Atominstitut,  Technische  Universit{\"a}t  Wien,  1020  Vienna,  Austria}
\affiliation{JST, PRESTO, 4--1--8 Honcho, Kawaguchi, Saitama, 332--0012, Japan}

\collaboration{Authors are listed in alphabetical order}

\begin{abstract}
Quantum error correction (QEC) is one of the central concepts in quantum information science and also has wide applications in fundamental physics. The \emph{capacity theorems} provide solid foundations of QEC\@. We here provide a general and highly applicable form of capacity theorem for both classical and quantum information, i.e., \emph{hybrid} information, with assistance of a limited resource of entanglement in one-shot scenario, which covers broader situations than the existing ones.
Harnessing the wide applicability of the theorem, we show that a demonstration of QEC by short random quantum circuits is feasible and that QEC is intrinsic in quantum chaotic systems. Our results bridge the progress in quantum information theory, near-future quantum technology, and fundamental physics.
\end{abstract}

\maketitle%

\section{Introduction}
Quantum error correction (QEC), a method of protecting information from quantum noise, is one of the central concepts in quantum information processing~\cite{S1995,S1996,CS1996}.
Since quantum systems are inevitably noisy due to uncontrollable interactions with environment, QEC has a wide range of applications in quantum communication, cryptography, and computation. In recent years, QEC has also shed new light on fundamental physics, providing a viewpoint to better understand quantum many-body phenomena such as topological orders~\cite{K2003,DKLP2002,K2006}, the black-hole information paradox~\cite{HP2007,YK2017,HP2019}, and a possible duality between quantum chaos and quantum gravity~\cite{M1999,W1998,TR2006,K2015,K2015Feb,PYHP2015,MS2016}. 

One of the central questions about QEC is how much information can \emph{in principle} be protected from a given noise. Since any quantum noise is formulated by a quantum channel, \emph{quantum channel capacity theorems} answer to this question. Depending on the type of information to be protected, either quantum or classical, and on resources available such as entanglement, numerous studies have been done~\cite{holevo98,schumacher97,lloyd1997capacity,devetak2005private,shor2002quantum,devetak2005capacity,bennett1999entanglement,devetak2004family}. For the asymptotic scenario of infinitely many uses of a noisy quantum channel, these results are merged into a unified formula in Ref.~\cite{hsieh2010entanglement}.

The asymptotic results are, however, applicable only when encoding and decoding can be applied coherently on a huge number of qubits, resulting in a difficulty of experimental demonstrations and of practical applications to fundamental physics.
In contrast, recent studies develop analyses without taking the asymptotic limit, which is called the \emph{one-shot} scenarios~\cite{dupuis2014decoupling,datta2011apex,buscemi2010quantum,datta2012one,mosonyi2009generalized,wang2012one,datta2013smooth,matthews2014finite,salek2019one,qi2018applications,datta2016second,anshu2018building,matthews2014finite,anshu2019near,salek2018one,wilde2017position,renes2011noisy,radhakrishnan2017one}. Since one-shot analyses overcome the difficulties of the asymptotic results~\cite{Y}, they are of great importance both theoretically and experimentally.

Despite the fact that the one-shot scenario is more practical,
little was explored about explicit encodings in one-shot scenario.
Moreover, the analyses in the one-shot scenario have been based on a rather specific technique, such as decoupling~\cite{dupuis2014decoupling,datta2011apex,buscemi2010quantum,datta2012one} and hypothesis testing~\cite{mosonyi2009generalized,wang2012one,datta2013smooth,matthews2014finite,salek2019one,qi2018applications,datta2016second,anshu2018building,matthews2014finite,anshu2019near,salek2018one,wilde2017position}, making it challenging to deal with all situations in a unified framework. A unified framework is helpful in exploring more applications of QEC not only in quantum information but also in fundamental physics.

In this paper, we provide a unified one-shot quantum capacity theorem for \emph{hybrid} information of classical and quantum with assistance of a limited amount of entanglement. The theorem covers broader situations than ever, and existing capacity theorem~\cite{holevo98,schumacher97,lloyd1997capacity,devetak2005private,shor2002quantum,devetak2005capacity,bennett1999entanglement,devetak2004family,hsieh2010entanglement,dupuis2014decoupling,datta2011apex,buscemi2010quantum,datta2012one,mosonyi2009generalized,wang2012one,datta2013smooth,matthews2014finite,salek2019one,qi2018applications,datta2016second,anshu2018building,matthews2014finite,anshu2019near,salek2018one,wilde2017position,renes2011noisy} are directly obtained as corollaries.
Our theorem is given in a concise manner, so that it can be directly applied to various problems. 
In particular, we provide two applications: one is entanglement-assisted quantum error correction (EAQEC)~\cite{BDH2006,HDB2007,LB2013} using short random quantum circuits (RQCs), and the other is QEC in quantum chaotic systems~\cite{CBQA2020, GH2020}.
In the former case, we show that RQCs with sublinear depth on a few tens of qubits can be used as a good encoder of hybrid information, even if they are noisy such as the gate fidelity around $99.5\%$. This implies that a proof-of-principle demonstration of EAQEC by near-future noisy intermediate-scale quantum (NISQ) devices~\cite{Preskill2018quantumcomputingin} is within reach of current quantum technology. 
In the latter, we quantitatively provide supporting evidences of the conjecture in strongly-correlated many-body physics that QEC is intrinsic in quantum chaos. Since QEC is also the key in the duality between quantum chaos and gravity, this contributes to better and quantitative understanding of the exotic physics between them.

This paper is organized as follows. We start with preliminaries in Sec.~\ref{S:pre}.
Our main result about the one-shot capacity theorem for hybrid information is summarized in Sec.~\ref{S:Cap}. Applications of the capacity theorem are provided in Sec.~\ref{S:Application}. After we summarize the paper in Sec.~\ref{S:conc}, we show technical details in Appendices.

\section{Preliminaries} \label{S:pre}
We start with introducing our notation in Subsec.~\ref{SS:notation}. The setting of hybrid information is explained in Subsec.~\ref{SS:setting}.

\subsection{Notations} \label{SS:notation}
We use superscripts to represent the systems on which operators and maps are defined, such as $\rho^{AB}$ for an operator on $AB$, $\mc{N}^{A \rightarrow B}$ for a superoperator from $A$ to $B$, and so on. 
A reduced operator on $A$ of $\rho^{AB}$ is denoted by $\rho^A$, i.e., $\rho^A = \tr_B [\rho^{AB}]$. 
As the identity operator $I$ acts trivially, $(M^{A} \otimes I^B )\rho^{AB} (M^{A } \otimes I^B )^\dagger$  is denoted simply by $M^A\rho^{AB} M^{\dagger A }$.
This notation is also used for superoperators: $\cN^{A \rightarrow C} \otimes {\rm id}^B(\rho^{AB})$ is denoted by $\cN^{A \rightarrow C} (\rho^{AB})$.
With slight abuse of notation, a system composed of $N$ duplicates of a system $A$ is denoted by $A^N$.

A maximally entangled state between $A$ and $B$ with the Schmidt rank $2^r$ is denoted by $\Phi_r^{AB}$. The completely mixed state on $A$ with rank $2^r$ is denoted by $\pi^A_r$. Using the above notation of a reduced state, we have $\Phi^A_r = \pi^A_r$.

The trace norm is defined by $\| X \|_1 \coloneqq \tr \sqrt{X X^{\dagger}}$. The purified distance $P(\rho, \sigma)$ between two subnormalized  positive semidefinite operators $\rho$ and $\sigma$ is defined by
\begin{equation}
P(\rho, \sigma) \coloneqq \sqrt{1 - \bar{F}(\rho, \sigma)^2},
\end{equation}
where $\bar{F}(\rho, \sigma) \coloneqq |\!| \sqrt{\rho} \sqrt{\sigma} |\!|_1  + \sqrt{(1- \tr [\rho])(1- \tr [\sigma])}$ is the purified fidelity~\cite{T16}.

The conditional max-entropy is defined for a state $\rho^{AB}$ by
\begin{align}
&H_{\rm max}(A|B)_{\rho} = \sup_{\varphi}\log_2 \|\sqrt{\rho^{AB}}\sqrt{I^A \otimes \varphi^B}\|_1^2,
\end{align}
where the supremum is taken over all normalized state $\varphi$ on the system $B$.
The smooth conditional max-entropy $H_{\rm max}^\epsilon(A|B)_{\rho}$ for a state $\rho$ on $AB$ and $\epsilon >0$ is defined by
\begin{equation}
H_{\rm max}^\epsilon(A|B)_{\rho} \coloneqq \inf_{\tilde{\rho}}H_{\rm max}(A|B)_{\tilde{\rho}},
\end{equation}
where the infimum is taken over all subnormalized positive semidefinite operators $\tilde{\rho}$ that are $\epsilon$-close to $\rho$ in the sense that $P(\rho, \tilde{\rho}) \leq \epsilon$. 
See Ref.~\cite{T16} for an introduction of entropies.

We use the Choi-Jamio\l kowski representation~\cite{J1972,C1975}. 

\begin{Lemma} \label{Lemma:CJ}
The Choi-Jamio\l kowski representation of a superoperator $\cT^{A \rightarrow B}$ from $A$ to $B$ is given by an operator $J(\mc{T}^{A \rightarrow B})$ on $AB$ defined by 
\begin{equation}
J(\mc{T}^{A \rightarrow B}) := ({\rm id}^A \otimes \mc{T}^{A' \rightarrow B})(\Phi^{AA'}),
\end{equation}
where ${\rm id}^A$ is the identity map on $A$, $\Phi^{AA'}$ is the maximally entangled state between $A A'$, and $A'$ is isomorphic to $A$.
The map $J$ is an isomorphism, and the inverse map $J^{-1}$ takes an operator $\rho^{AB}$ on $AB$ to the superoperator $\mc{T}^{A \rightarrow B}$ given by
\begin{equation}
\mc{T}^{A\rightarrow B}(\sigma^A) = d_A \tr_A \bigl[ \bigl( (\sigma^A)^T \otimes I^B \bigr) \rho_{AB} \bigr]
\end{equation}
for any $\sigma_A$ on $A$, where $T$ represents a transpose with respect to the Schmidt basis of $\Phi^{AA'}$.
\end{Lemma}

The Choi-Jamil\l kowski representation provides an isomorphism between a set of all CPTP maps from $A$ to $B$ and a set of states on $AB$ such that its marginal state on $A$ is the completely mixed state. This isomorphism is also known as the channel-state duality.

\subsection{Setting of hybrid channel coding} \label{SS:setting}

Let $\mathcal{N}^{A \rightarrow B}$ be a noisy quantum channel with an input $A$ and an output $B$, represented by a completely positive (CP) and trace-preserving (TP) map. 
We consider a task to encode $C$-bit classical and $Q$-qubit quantum information, which we call \emph{hybrid} information, to system $A$ in such a way that it is protected from the noise $\mathcal{N}^{A \rightarrow B}$ as much as possible. Entanglement of $E$ ebits can be used as extra resource. Hence, the situation is entanglement-assisted quantum error correction (EAQEC).
Note that the channel $\mathcal{N}$ is not necessarily a communication channel between distant places but can be a noise on a physical system onto which information will be stored. 

The task is formally described as follows. Let $M$ be the hybrid information source that is split into a classical part $M_c$ of $C$ bits and a quantum part $M_q$ of $Q$ qubits. 
We use a uniform probability distribution $p(j)=2^{-C}$ over $j \in \{0, \dots, 2^C-1\}$ and a maximally entangled state $\Phi^{M_q R}_Q$ of $Q$ ebits between $M_q$ and a reference $R_q$. 
The encoding operation is represented by a family of CPTP maps $\{\tilde{\mathcal{E}}_j^{M_q F_A \rightarrow A}\}_j$, and the decoding operation by an instrument $\{\tilde{\mathcal{D}}_j^{B F_B \rightarrow M_q}\}_j$, i.e., a family of CP maps that sum up to a TP map.
We say that $C$-bit classical and $Q$-qubit quantum information is protected from a noise $\cN$ using $E$ ebits within an error $\delta >0$, or simply {\it the tuple $(C,Q,E,\delta)$ is achievable for $\cN$},
if there exist $\{ \tilde{\mathcal{E}}_j \}_j$ and $\{ \tilde{\mathcal{D}}_j \}_j$ such that
\begin{equation}
\frac{1}{2^{C}} \sum_j |\!| \tilde{\mathcal{D}}_{j} \circ \mathcal{N} \circ \tilde{\mathcal{E}}_j(\Phi_{Q}^{M_q R_q} \otimes \Phi_E^{F_AF_B}) - \Phi_{Q}^{M_q R_q} |\!|_1 \leq \delta, \label{Task1}
\end{equation}
where $\Phi^{F_A F_B}_{E}$ denotes the $E$ ebits.
Equation~(\ref{Task1}) implies that the initial symbol $j$ of the classical information and the initial state $\ket{\psi}$ of the quantum information can be recovered within error $\delta$ on average after the decoding operation.
Thus, if there is a tuple $(C,Q,E,\delta)$ with sufficiently small $\delta$, then EAQEC for the $C$-bit and $Q$-qubit informations is possible.

The achievability condition~\eqref{Task1} can be equivalently rephrased in terms of a \emph{hybrid source state}, which is more convenient in the analysis in Subsec.~\ref{SS:Application1}.
Using a fixed orthogonal basis $\{ \ket{j} \}_{j}^{2^C}$ in $M_c$ and $R_c$, we define the hybrid source state by 
\begin{equation}
\Phi_{C,Q}^{MR}= \Omega_C^{M_c R_c} \otimes \Phi_{Q}^{M_qR_q}, \label{Eq:m}
\end{equation}
where 
\begin{equation}
\Omega_C^{M_cR_c} \coloneqq 
\frac{1}{2^C}\sum_{j=1}^{2^C}\proj{j}^{M_c} \otm\proj{j}^{R_c}
\label{omegaCdfn}
\end{equation}
is a classically correlated state.
Based on the hybrid source state, ~\eqref{Task1} is equivalent to the condition that
\begin{equation}
\frac{1}{2}
\bigl|\! \bigl| \tilde{\mathcal{D}} \circ \mathcal{N} \circ \tilde{\mathcal{E}}(\Phi_{C,Q}^{M R} \otimes \Phi_E^{F_AF_B}) - \Phi_{C,Q}^{M R} \bigr|\! \bigr|_1 \leq \delta. \label{Eq:fin}
\end{equation}
Here, $\tilde{\mathcal{E}}^{M F_A \rightarrow A}$ and  $\tilde{\mathcal{D}}^{B F_B \rightarrow M}$ are CPTP maps that represent the encoding and decoding operations, respectively.
Those are related with $\{\tilde{\mathcal{E}}_j^{M_qF_A\rightarrow A}\}_j$ and $\{\tilde{\mathcal{D}}_j^{BF_B\rightarrow M_q}\}_j$ by
\begin{align}
\tilde{\cE}^{MF_A \rightarrow A} (\rho) &\coloneqq \sum_j \tilde{\cE}_j^{M_q F_A \rightarrow A}( \bra{j}^{M_c} \rho^{MF_A} \ket{j}^{M_c} ),\\
\tilde{\cD}^{BF_B \rightarrow M} (\sigma) &\coloneqq \sum_j \ketbra{j}{j}^{M_c} \otimes \tilde{\cD}_j^{M_q F_A \rightarrow M_q}( \sigma^{M_q F_A}),
\end{align}
which are equivalent to
\begin{align}
\tilde{\cE}_j^{M_q F_A \rightarrow A}(\rho') &\coloneqq \tilde{\cE}^{MF_A \rightarrow A}(\ketbra{j}{j}^{M_c}  \otimes \rho'),\label{eq:cEcEj}\\
\tilde{\cD}_j^{BF_B \rightarrow M_q} (\sigma') &\coloneqq \tr_{M_c} [\ketbra{j}{j}^{M_c} \tilde{\cD}^{BF_B \rightarrow M}(\sigma')],\label{eq:cDcDj}
\end{align}
respectively. See Appendix~\ref{App:EquivCond} for the proof.

\section{One-Shot Capacity Theorem for Hybrid Information} \label{S:Cap}

In this section, we state our main result about a one-shot capacity theorem for hybrid information. We start with a direct part, which provides a tuple $(C,Q,E,\delta)$ achievable by a specific encoding scheme, and then provide its converse, which shows that no encoding scheme can substantially improve the achievable tuple.

\begin{figure}[t!]
\centering
\includegraphics[width=80mm,clip]{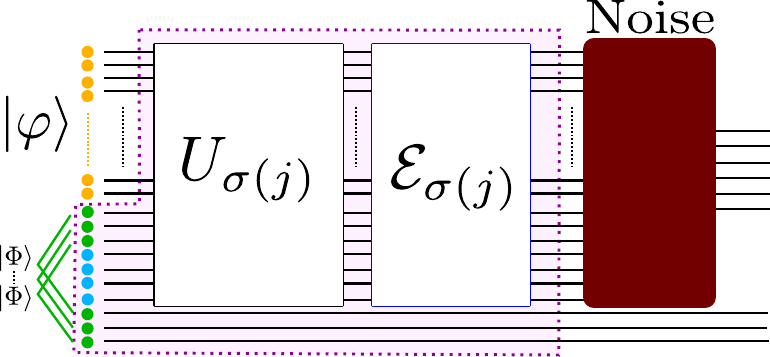}
\caption{A diagram of the encoding scheme used in Theorem~\ref{Thm:achiSM}. The circles in the left represent qubits. Yellow ones are those carrying quantum information, green ones are entangled qubits, and blue ones are ancillary qubits that are initially set to any pure state, such as $\ket{0}^{\otimes m}$, where $m$ is the number of the ancillary qubits.}
\label{Fig1}
\end{figure}

To state the direct part, we consider an encoding scheme that is composed of classical preprocessing and quantum encoding, which will turn out to be nearly optimal. In the classical preprocessing, we embed the alphabet $\{0, \dots, 2^C-1 \}$ to a larger one $\{0, \dots, J-1\}$, where $J \geq 2^C$, and apply a random permutation $\sigma$, yielding $\sigma(j)$. We then encode the information into a quantum system by three steps (see also Fig.~\ref{Fig1}):
\begin{enumerate}
\item Ancillary qubits are attached to enlarge the composite system $M_q F_A$ to a larger system labeled as $S_r$.
\item A random unitary $U_{\sigma(j)}$ is applied onto $S_r$, where $U_{j'}$ for each $j'$ is sampled from the unitary group uniformly and independently at random with respect to the Haar measure~\cite{DupuisThesis}. Such a unitary is called a Haar random unitary.
\item A CPTP map $\cE_{\sigma(j)}:S_r\to A$ is applied.
\end{enumerate}
Note that a Haar random unitary in the second step and the CPTP map in the third step are chosen depending on the outcome $\sigma(j)$ of the classical preprocessing. The whole encoding operation is given by 
\begin{equation}
    \tilde{\ca{E}}_j:=\ca{E}_{\sigma(j)}\circ U_{\sigma(j)}\circ \Gamma,
\end{equation} 
with $\Gamma:M_qF_A\rightarrow S_r$ representing the attachment of ancillary qubits. This scheme forms a family of encoding operations specified by $\{\cE_j\}_{j = 0}^{J-1}$ in the third step. 

To characterize the tuple $(C,Q,E,\delta)$ achievable by this encoding scheme, it is convenient to use the channel-state duality (see Lemma~\ref{Lemma:CJ}) to represent each $\cE_j$ by a state $\rho_j := J(\cE_j)$ on $S_r A$. Letting $S_c$ be a system with a fixed basis $\{ \ket{j} \}_{j=0}^{J-1}$, and $S=S_cS_r$, we define a state $\rho$ on $SA$ by
\begin{equation}
\rho^{SA} \coloneqq \frac{1}{J} \sum_{j=1}^J \proj{j}^{S_c} \otimes \rho_j^{S_rA}, \label{Eq:rho}
\end{equation}
with the property that the marginal state $\rho^S$ is the completely mixed state. The family $\{\cE_j\}_{j=0}^{J-1}$ of encoding operations is fully specified by this state $\rho$.

We now state our first result that the achievable tuple $(C,Q,E,\delta)$ for $\cN$ is characterized by the smooth conditional max-entropy $H_{\rm max}^{\epsilon}$ of the state $\rho_{\cN} \coloneqq ( {\rm id}^S \otimes \cN^{A \rightarrow B})(\rho^{SA})$.

\begin{thm} \label{Thm:achiSM}(Direct part)
Let $\epsilon, \delta_1, \delta_2 > 0$, and $d_{S_r}$ be the dimension of the Hilbert space of $S_r$. If $\rho_{\cN}$ and $(C,Q,E)$ $(C,Q \neq 0)$ satisfy
\begin{align}
Q\!+\!E & \leq \log_2 d_{S_r}, \label{Ineq:a1} \\
C\!+\!Q\!-\!E& \leq -H_{\rm max}^{\epsilon}(S|B)_{\rho_{\cN}}+\log_2{(J\!-\!1)} +\log_2{\delta_1}, \!\!\label{Ineq:a2}\\ 
Q\!-\!E & \leq -H_{\rm max}^{\epsilon}(S_r|BS_c)_{\rho_{\cN}} +\log_2{\delta_2}, \label{Ineq:a3}
\end{align}
the encoding scheme described above achieves $(C,Q,E, \delta)$ with $\delta \leq \sqrt{\sqrt{\delta_1}+\sqrt{\delta_2}+4\epsilon}$, for almost any choice of random permutations $\sigma$ and random unitaries $\{U_j\}_j$.
If $C=0$, the same statement holds by removing the condition~\eqref{Ineq:a2} and setting $\delta_1=0$.
\end{thm}

Conversely, we can also show that the achievable tuple in Theorem~\ref{Thm:achiSM} cannot be significantly outperformed by \emph{any} encoding map.

\begin{thm}\label{Thm:convSM}(Converse part)
Suppose that a tuple $(C,Q,E,\delta)$ is achievable for $\ca{N}$. 
Regardless of the encoding scheme, there exist $J$- and $d_r$-dimensional systems $S_c$ and $S_r$, respectively, and a state $\rho$ in the form of
\begin{equation}
\rho^{SA} = \frac{1}{J} \sum_{j=1}^J \proj{j}^{S_c} \otimes \rho_j^{S_rA}, \label{Eq:rho}
\end{equation}
whose marginal state $\rho^S$ on $S$ is the completely mixed state, such that for any $\iota\in(0,1]$,
\begin{align}
Q\!+\!E& \leq \log_2 d_{S_r}, \label{Ineq:c1}\\
C\!+\!Q\!-\!E&\leq -H_{\rm max}^{\lambda}(S|B)_{\rho_{\cN}}+\log_2{J}-\log_2{\iota}, \label{Ineq:c2}\\
Q\!-\!E & \leq -H_{\rm max}^{\lambda'}(S_r|BS_c)_{\rho_{\cN}}-\log_2{\iota}, \label{Ineq:c3}
\end{align}
where $\lambda$ and $\lambda'$ depend on $\iota$ and $\delta$ and vanish as $\iota, \delta\rightarrow0$.
\end{thm}

Since~\eqref{Ineq:c1},~\eqref{Ineq:c2} and~\eqref{Ineq:c3} in Theorem~\ref{Thm:convSM} coincide~\eqref{Ineq:a1},~\eqref{Ineq:a2} and~\eqref{Ineq:a3} in Theorem~\ref{Thm:achiSM} up to a change of smoothing parameters, the encoding scheme used in Theorem~\ref{Thm:achiSM} is nearly optimal.

Theorems~\ref{Thm:achiSM} and~\ref{Thm:convSM} reveal how much information is protected from a given noise with an explicit encoding scheme, where the states $\{\rho_j\}$, specifying the CPTP maps $\{\cE_j\}$ used in the encoding scheme, are treated as parameters. This formulation is useful, given that not all maps are implementable by current noisy quantum devices: by substituting the map that is realizable in an experimental system into Theorem~\ref{Thm:achiSM}, we can reveal how much information can be protected from a given noise in that system. This leads us to explore the possibility to demonstrating QEC as shown later.

The full achievable rate region can be obtained from Theorem~\ref{Thm:achiSM} by taking the union of $(C,Q,E,\delta)$ over all possible choices of CPTP maps $\{\ca{E}_j\}$, or equivalently, over all states $\rho$ in the form of (\ref{Eq:rho}).
Important capacity theorems such as the Holevo-Schumacher-Westmoreland theorem~\cite{holevo98,schumacher97} for classical information, the Lloyd-Shor-Devetak theorem~\cite{lloyd1997capacity,shor2002quantum,devetak2005private} for quantum information, Devetak-Shor theorem~\cite{devetak2005capacity} for hybrid information, those with assistance of entanglement~\cite{bennett1999entanglement,devetak2004family}, and their extensions to the one-shot scenario~\cite{dupuis2014decoupling,datta2011apex,buscemi2010quantum,datta2012one,mosonyi2009generalized,wang2012one,datta2013smooth,matthews2014finite,salek2019one,qi2018applications,datta2016second,anshu2018building,matthews2014finite,anshu2019near,salek2018one,wilde2017position,renes2011noisy,radhakrishnan2017one}, readily follow from our result. Thus, our results interpolate these theorems and lead to the full characterization of classical and quantum capacities in the one-shot scenario. See Ref.~\cite{wakakuwa2020randomized} for the explicit reductions.

The proofs of Theorems~\ref{Thm:achiSM} and~\ref{Thm:convSM} are based on the observation that achieving the task presented above is equivalent to achieving the \emph{partial decoupling}~\cite{wakakuwa2019one} and are given in~\cite{wakakuwa2020randomized}.
Due to (\ref{Eq:rho}), the state $\Psi$ obtained from $\ca{N}\circ \ca{E}$ by the channel-state duality is in the form of $\Psi^{S\bar{E}}=\sum_{j,k}\proj{jk}^{S_c}\otm\Psi_{jk}^{S_r\bar{E}_r}\otm\proj{jk}^{E_c}$. 
The randomized partial decoupling theorem~\cite{wakakuwa2019one} deals with a situation in which a state in this form is transformed by a random permutation $\sigma$ on $S_c$ and a random unitary in the form of $\sum_j\proj{j}^{S_c}\otm U_j^{S_r}$ followed by a CP map $\ca{T}$ on $S$.  
We particularly choose $\ca{T}$ to be a projection onto a subspace followed by a partial trace, the dimensions of which are given by $c$, $q$ and $e$.
The condition for successfully achieving the partial decoupling is represented in terms of the entropies of the state $\rho_\ca{N}$, which yields Eqs.~\eqref{Ineq:a1}-\eqref{Ineq:a3}. 
This is similar to the standard technique in the derivation of quantum capacity from decoupling~\cite{ADHW2009}.
Theorem~\ref{Thm:convSM} also follows from the converse bound of partial decoupling.

\section{Applications} \label{S:Application}

The one-shot capacity theorem, i.e. Theorems~\ref{Thm:achiSM} and~\ref{Thm:convSM}, has far-reaching consequences beyond quantum communication since it provides the fundamental limit of EAQEC. To demonstrate its potential uses in various situations, we consider the simplest encoding scheme that Theorem~\ref{Thm:achiSM} can deal with. That is, we consider the encodings, whose quantum part consists only of attaching ancillary qubits and unitary operations. Based on this, we provide two applications. One is EAQEC by short RQCs and the other is QEC in quantum chaotic systems. 

In Subsec.~\ref{SS:s}, we explain the setting common in both applications. We address EAQEC by short RQCs in Subsec.~\ref{SS:Application1} and QEC in quantum chotic systems in Subsec.~\ref{SS:Application2}.

\subsection{Setting --QEC by unitary dynamics--} \label{SS:s} 

We consider the task to store $(cN)$-classical and $(q N)$-quantum information in $N:=m n$ qubits, where each $m$ qubits are exposed to the noise independently, with assistance of arbitrary amount of entanglement.
Unlike the encoding used in Theorem~\ref{Thm:achiSM}, we assume that the quantum part of encoding is limited to using ancillary qubits and applying a unitary operation. That is, we assume that $\ca{E}_j$ in Theorem~\ref{Thm:achiSM} is the identity map.
For the error suppression to be possible, we also assume that $c \leq 2$ and $q \leq 1$. 

From Theorem~\ref{Thm:achiSM}, an upper bound on the error achievable by this encoding scheme is immediately obtained.
For a given set of $m$-qubit noises $\{ \cN_i \}_i$ ($i=1, \dots, n$), we denote by $\delta(c, q|\cN, n)$ the recovery error of the $(cN)$-bit and $(qN)$-qubit information, where
\begin{equation}
    \cN = \bigotimes_{i=1}^n \cN_i.
\end{equation}
Let $A$ be the $N$-qubit system on which the unitary $U_{\sigma(j)}$ are applied, and $a$ be the $m$-qubit system on which the noisy channel acts. Note that $\cH^A = (\cH^a)^{\otimes n}$. We also introduce a virtual system $A'$ and $a'$, whose Hilbert space is isomorphic to that of $A$ and $a$, respectively.
We apply Theorem~\ref{Thm:achiSM} under the following correspondence:
\begin{align}
&
S_r \rightarrow A,
\quad
B \rightarrow A',
\\
& C\rightarrow cN,
\quad
Q\rightarrow qN,
\quad
E\rightarrow (1-q)N,
\\
&
d_{S_r}\rightarrow 2^n,
\quad
\epsilon\rightarrow 0.
\end{align}
We obtain that, if the following holds,
\begin{align}
(c+2q-1)N & \leq -H_{\rm max}(S_c A |A')_{\rho_{\cN}} \notag \\
&\hspace{15mm}+\log_2{(J-1)} +\log_2{\delta_1}, \label{Ineq:1}\\
(2q-1) N & \leq -H_{\rm max}(A |A' S_c)_{\rho_{\cN}} +\log_2{\delta_2}, \label{Ineq:2}
\end{align}
then $\delta(c, q|\cN, N) \leq \sqrt{\sqrt{\delta_1} +\sqrt{\delta_2}}$.

Since $\cE_j$ is the identity map for all $j$ in this situation, we have $\rho_j^{A A'} = \Phi_{N}^{A A'} = (\Phi_m^{aa'})^{\otimes n}$.
Thus, $\rho_{\cN}$ is explicitly given by
\begin{equation}
\rho_{\cN}^{S_c A A'} = \pi_{\log_2 J}^{S_c}  \otimes_{i=1}^n \cN_i^{a'}(\Phi_m^{aa'}).
\end{equation}
It is straightforward to compute the conditional max-entropies of this state by using the additivity for the tensor product:
\begin{align}
H_{\rm max}(S_c A |A')_{\rho_{\cN}} &= \log_2 J + n h_{\cN},\\
H_{\rm max}(A |A' S_c)_{\rho_{\cN}} &=n h_{\cN},
\end{align}
where 
\begin{equation}
    h_{\cN} = \frac{1}{n} \sum_{i=1}^n H_{\rm max}(a|a')_{\Phi_{\cN_i}}, \label{Eq:CondMax}
\end{equation}
with $\Phi_{\cN_i}^{aa'}$ being $({\rm id}^a \otimes \cN^{a'}_i)(\Phi^{aa'}_m)$. Note that $-m \leq h_{\cN} \leq m$.

By taking the minimum values of $\delta_1$ and $\delta_2$ that satisfy Eqs.~\eqref{Ineq:1} and~\eqref{Ineq:2}, respectively, we arrive at
\begin{align}
&\delta_1 = \left( 1+ \frac{1}{J-1} \right) 2^{(c+2q + \frac{h_{\cN}}{m} - 1)N},\\
&\delta_2 = 2^{(2q + \frac{h_{\cN}}{m} - 1)N}.
\end{align}
Since $J$ can be chosen arbitrarily large in the preprocessing, we assume that $1/(J-1) \approx 0$. 
By separately considering the case of $c=0$, we arrive at
\begin{equation}
\delta(c, q|\cN, n)
\lesssim
\begin{cases}
\bigl( 1+ 2^{ \frac{cN}{2}} \bigr)^{1/2} 2^{(2q +  \frac{h_{\cN}}{m} - 1)\frac{N}{4},} & (c \neq 0), \\
2^{(2q +  \frac{h_{\cN}}{m} - 1)\frac{N}{4}}, & (c = 0).
\end{cases} \label{Ineq:error}
\end{equation}
This is the basic formula that we use in the following applications.

\subsection{Application 1 --EAQEC by short RQCs--} \label{SS:Application1}

\begin{figure}[t!]
\centering
\includegraphics[width=80mm,clip]{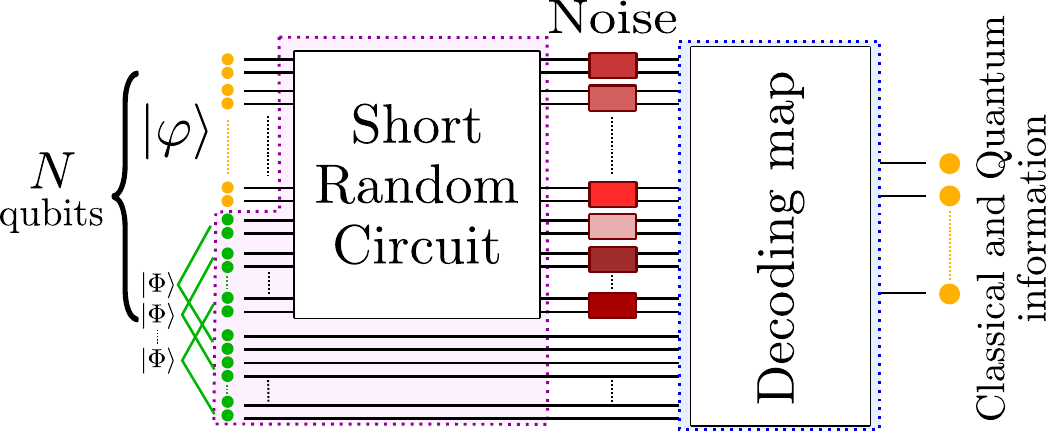}
\caption{A diagram of EAQEC we consider in the first application. The red part corresponds to a quantum part of the encoding, i.e., attaching $(1-q)N$ ancillary qubits and applying short random circuits depending on the outcome $\sigma(j)$ of the classical preprocessing. The ancillary qubits are all entangled with other $(1-q)N$ qubits that are used in the decoding operation. The noise is depicted by the boxes of various red colors. Each box acts on $m$ qubits, and different colors indicate that they may be different noises. After the noise, a decoding map is applied, which is indicated by a blue box in the diagram, so as to extract $(cN)$-bit and $(qN)$-qubit information.}
\label{Fig2}
\end{figure}

As the first application of \eqref{Ineq:error}, we show that one-shot EAQEC~\cite{BDH2006,HDB2007,LB2013} of independent single-qubit noises is feasible by short random quantum circuits (RQCs) even when they are noisy. RQCs are the circuits, in which randomly chosen gates are applied to nearest-neighbor qubits at each step, and have used to experimentally demonstrate quantum supremacy~\cite{G2019}. As depicted in Fig.~\ref{Fig2}, we consider the situation where the unitary encoding part is given by RQCs with shallow depth.

The key observation is that RQCs with nearest-neighbor gates on a 2-dimensional lattice have the same second-order statistical moments of a Haar random unitary, if the depth is $O(\sqrt{N})$~\cite{HM2018}. It is known that this property suffices to reproduce Theorem~\ref{Thm:achiSM}. Hence, even when the unitary part of the encoding is replaced with shallow RQCs, we can directly apply the formula~\eqref{Ineq:error} and can reveal the achievable errors.

To apply~\eqref{Ineq:error}, we need to compute the conditional max-entropy $h_{\cN}$, defined by~\eqref{Eq:CondMax}, for a given noise. In the following analyses, we use the expression given in Ref.~\cite{V} to represent the conditional max-entropy $H_{\rm max}(\cdot|\cdot)$ in terms of a solution of semidefinite programming (SDP). By numerically solving this SDP using YALMIP~\cite{L} and Splitting Conic Solver (SCS)~\cite{O}, we evaluate $h_{\cN}$.

We especially deal with two types of one-qubit noise ($m=1$), i.e., dephasing $\mathcal{D}_{p}$ and amplitude damping $\cA_{\gamma}$ noises~\cite{P}:
\begin{align}
    &\mathcal{D}_{p} (\rho ) = \left(1-\frac{p}{2}\right) \rho + \frac{p}{2} Z \rho Z,\laeq{dfnDp}\\
    &\mathcal{A}_{\gamma} (\rho ) = K_0 \rho K_0^{\dagger} + K_1 \rho K_1^{\dagger},
\end{align}
where $p$ and $\gamma$ are error parameters taking the value between $0$ and $1$, $Z$ is the Pauli $Z$ operator, $K_0 = |0 \rangle \langle 0| + \sqrt{1-\gamma} |1 \rangle \langle 1|$, and $K_1 = \sqrt{\gamma} |0 \rangle \langle 1|$. 
In both cases, we assume that the noise are independently acting on each qubit, implying that, for all $i$, $\cN_i = \mathcal{D}_{p}$ for the former and $\cN_i = \cA_{\gamma}$ for the latter.

Below, we consider two types of RQCs, one is noiseless and the other is noisy, and reveal the recovery errors that can be achieved by the encoding based on noiseless/noisy RQCs. We then comment on how to decode information.

\begin{figure}[t!]
\centering
\includegraphics[width=80mm,clip]{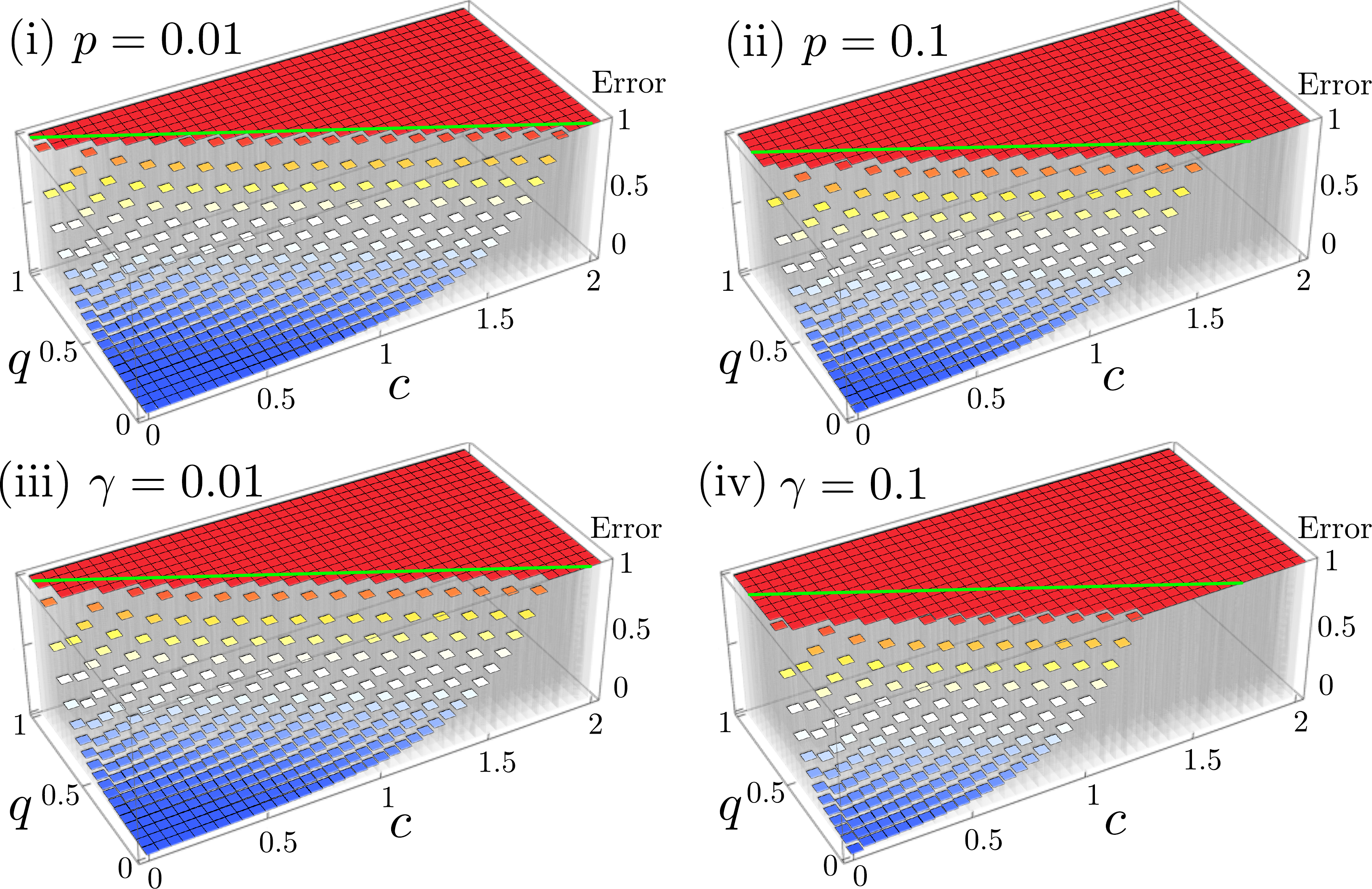}
\caption{Upper bounds on errors $\delta(c,q| \cN, N)$, where $N$ is set to $20$. The panels (i) and (ii) show the errors for the dephasing noise $\cD_p$ with $p= 0.01$ and $0.1$, respectively, while (iii) and (iv) show those for the amplitude damping noise $\cA_{\gamma}$with $\gamma= 0.01$ and $0.1$, respectively. For comparison, we also indicate the optimal line in the asymptotic case by green in each figure. If the pair $(c,q)$ is outside the green line, the error can never be significantly suppressed by any means in the limit of infinitely many uses of the noisy qubits.}
\label{fig:capacity}
\end{figure}

\subsubsection{Noiseless case}

From the numerically obtained values of $h_{\cN}$ for $\cN_i = \cD_p, \cA_{\gamma}$, the errors $\delta(c, q|\cD_p, N)$ and $\delta(c, q|\cA_{\gamma}, N)$ for several values of $p$ and $\gamma$ are given in Fig.~\ref{fig:capacity} for $N=20$. It is clear that the noises are suppressed for small $(c, q)$. We refer to the regions, where the noise is strictly smaller than $1$ (the region except the red panels), as the \emph{achievable region} by RQCs. 
In this region, the errors exponentially decrease as $N$ increases and, it is numerically confirmed that the errors become negligible when $N$ is moderately large, such as $n \approx 50$.

The green line in Fig.~\ref{fig:capacity} provides the optimal line in the asymptotic case: the errors for the $(c,q)$'s below the green line can be made arbitrarily small by a proper encoding when infinitely many noisy qubits are used, and any points above the line cannot. The optimal line was originally given in Ref.~\cite{hsieh2010entanglement} and also follows from Theorem~\ref{Thm:convSM}. 
From the figure, we observe that the region achievable by short RQCs is comparable to that below the green line, particularly for the dephasing noise with any $p$ and for the amplitude damping noise with small $\gamma$.
Since no encoding scheme can significantly suppress the error outside the asymptotically optimal line, this indicates that the performance of the encoder using short RQCs is close to that of the best ones.
On the other hand, a non-negligible gap remains in the case of $\gamma=0.1$, which is because the amplitude damping noise affects more on the subspace spanned by vectors with more $\ket{1}$s, while the encoder with short RQCs acts equally on the whole space.

\begin{figure}[t!]
\centering
\includegraphics[width=75mm,clip]{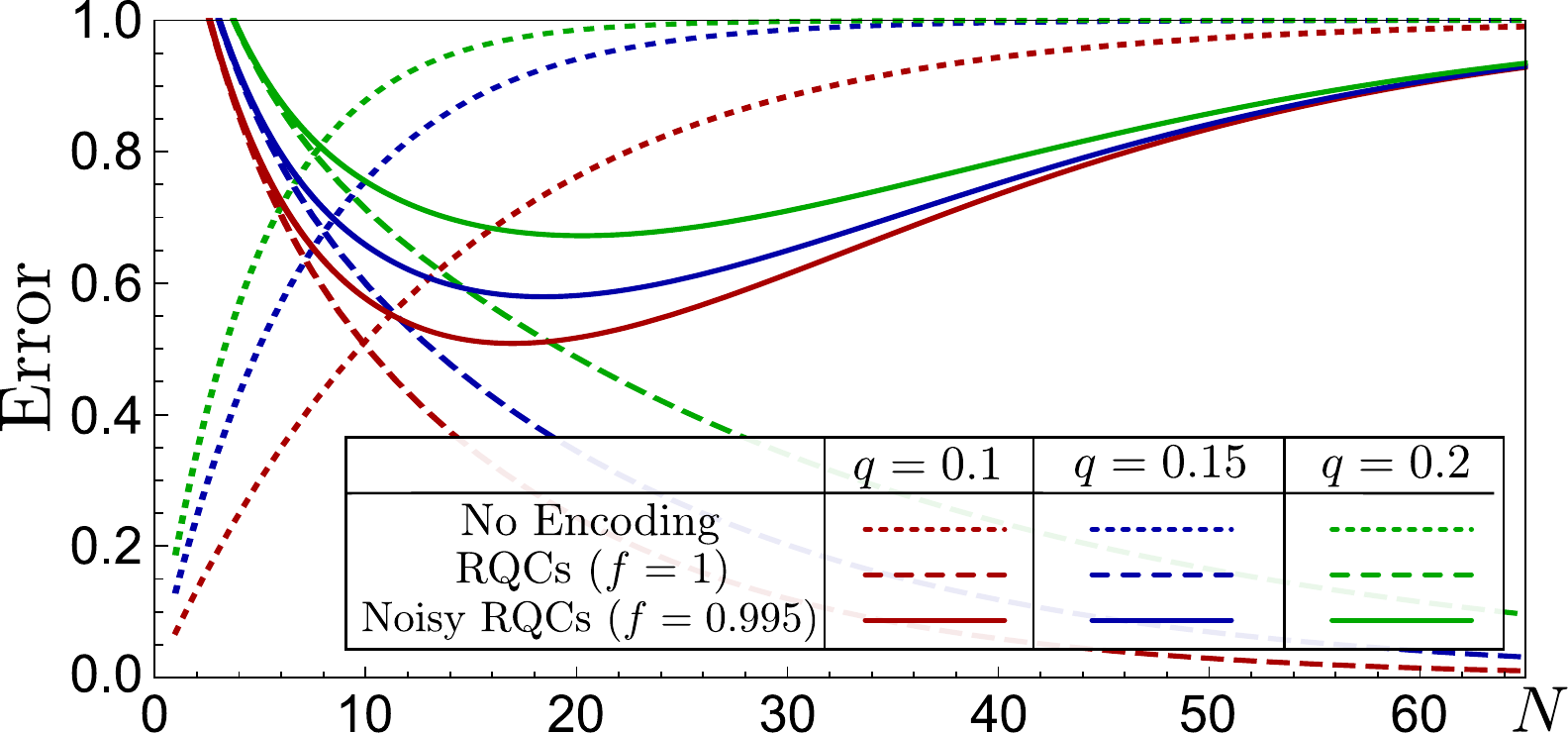}
\caption{Upper bounds on the errors $\delta(c,q|\cD_p, N)$ and $\delta_{\rm noisy}(c,q|\cD_p, N)$ for $p=5\%$ and $G = N^{3/2}$ as a function of the number $N$ of noisy qubits. The $c$ is fixed to $0.9$, while the $q$ varies from $0.1$ (red lines), $0.15$ (blue lines), and $0.2$ (green lines). The dotted lines correspond to the case (i) in the main text, the dashed lines plots $\delta(c,q|\cD_p, N)$ (case (ii)), and the solid lines $\delta_{\rm noisy}(c,q|\cD_p, N)$ with gate fidelity being $99.5\%$ (case (iii)).
}
\label{Fig:DP2D}
\end{figure}

\subsubsection{Noisy RQCs}

We next investigate the case when the short RQCs are noisy. Following Ref.~\cite{G2019}, we assume that the noise on each gate in the RQCs is depolarizing with gate fidelity $f = 99.5\%$, which is comparable with the reported value. This noise induces an additional error to~\eqref{Ineq:error}. With $G$ denoting the number of gates in the RQCs, the error can be evaluated by
\begin{equation}
\delta_{\rm noisy}(c,q| \cN, N, f) \lesssim (1-f^{G}) + f^G \delta(c,q| \cN, N).
\end{equation}

In Fig.~\ref{Fig:DP2D}, we provide the errors when (i) information is directly stored in noisy qubits, whose errors are investigated in Appnedix~\ref{app:EEWE}, (ii) noiseless short RQCs are used, and (iii) noisy short RQCs are used. Here, we mean by short that the number $G$ of gates in the circuit is $N^{3/2}$. 
The noiseless RQCs with $G = O(N^{3/2})$ form unitary $2$-designs~\cite{HM2018}. Shorter RQCs are also expected to achieve decoupling~\cite{BF2013,NHMW2017,GKHJF2020}. For simplicity, we here assume that the noiseless short RQCs with $G = N^{3/2}$ reproduce~\eqref{Ineq:error}.

By comparing (i) and (iii), we observe that, even by noisy short RQCs with a couple of dozens of qubits, noise suppression is possible to some extent in a certain parameter region, e.g., $c>0$, $q>0$, and $c+q\gtrsim 1$. Importantly, the demonstration becomes possible only when the hybrid information is to be stored. This indicates that short RQCs with moderately high gate fidelity can be used for the proof-of-principle demonstration of hybrid EAQEC\@.

\subsubsection{Decoding information}

So far, we have not explicitly considered how to decode the information. In general, explicit construction of a decoder for random encoding schemes is computationally intractable. In certain situations, however, it is possible to construct decoders efficiently. For instance, this is the case when the following two conditions are satisfied;
\begin{enumerate}
    \item each gate in the RQC is chosen from Clifford gates, namely, it is a random Clifford circuit;
    \item the noise is such that its Stinespring dilation is Clifford.
\end{enumerate}

In this situation, it is possible to classically simulate the encoding circuit and the Stinespring dilation of the noise in an efficient manner: since they are both Clifford, we can simply use the Gottesman-Knill theorem~\cite{PhysRevA.70.052328}. From the classical simulation of the encoding circuit and the Stinespring dilation of the noise, a decoding isometry can be constructed by a standard technique in the decoupling approach~\cite{DupuisThesis}. In this case, the decoding isometry is also Clifford. Hence, we can also implement the decoder efficiently.

One may think that the second assumption of the noise, i.e., its Stinespring dilation being Clifford, excludes a continuous parameter of the noise. This is circumvented by adding an extra assumption of the noise that the location of the qubits, on which the noise takes place, is available when the information is decoded. In this situation, the ratio of the total number of qubits and the number of noisy qubits represents a parameter of the noise, which can be arbitrarily close to be continuous by increasing the number of qubits. The erasure channel~\cite{PhysRevA.56.33,PhysRevLett.78.3217} is a canonical example of such a noise.

As our analysis suggests, noise suppression can be obtained using an RQC with depth $O(\sqrt{N})$ in a certain parameter region relevant to near-term experiments. The same degree of noise suppression is likely achievable by random Clifford circuits with the same depth~\cite{HMHEGR2020, HM2018}. Therefore, by implementing these encoder, noise, and decoder explicitly, the proof-of-principle demonstration of hybrid EAQEC is possible in the experiments.

\subsection{Application 2 --QEC and quantum chaos--} \label{SS:Application2}

As the second application, we consider QEC in quantum chaotic systems.
In the last decade, quantum chaos has been studied from quantum information-theoretic viewpoint based on out-of-time-ordered-correlators~\cite{RY2017} and operator entanglement~\cite{BL2005,HQRY2016}. The studies in the literature strongly indicate that chaotic dynamics is sufficiently information scrambling and has similar properties of a Haar random unitary~\cite{K2015,K2015Feb,MS2016,RD2015,HQRY2016,RY2017,GHST2020}. Based on the studies, the correspondence between quantum chaos and QEC has been often claimed. However, there has been no solid analysis about QEC in quantum chaotic systems. 
Here, by applying the formula~\eqref{Ineq:error}, we quantitatively show that spontaneous chaotic dynamics is a good encoder of quantum information, which results in the protection of information in the many-body system from additional noise induced by thermal environment. This will be the first quantitative supporting evidences of the conjecture that quantum chaos is QEC.

We especially consider the following thought experiment:
\begin{enumerate}
    \item Quantum information is initially stored in a subsystem of a chaotic system.
    \item The system undergoes the unitary time-evolution, which we assume to be information scrambling.
    \item The noise is induced into the system from an environment at finite temperature.
    \item The original quantum information is tried to be recovered from the noisy many-body system.
\end{enumerate} 
As an example of noise from the thermal environment, we especially consider that each pair $(i, i+1)$ of neighboring spin-$1/2$ particles in the system interacts with a common environmental spin-$1/2$ particle $E$ for time duration $t$. 

We assume that each environmental particle is initially
\begin{equation}
\rho_E(p) = p \ketbra{\! \downarrow}{\downarrow \!} + (1-p) \ketbra{\! \uparrow}{\uparrow \!},    
\end{equation}
where $p$ is a Boltzmann factor at the temperature of the environment.
In particular, when $p=1$, the environment is in a pure state, which corresponds to the case where the temperature is zero, and when $p=0.5$, the temperature is infinite.
We also assume that the interaction is given by an independent Heisenberg-type interaction with a fluctuating coupling constant, i.e.,
\begin{equation}
H_{i,i+1,E}(\vec{J}_{i,i+1}) := H_{i,E}(\vec{J}_i) + H_{i+1,E}(\vec{J}_{i+1}),
\end{equation}
where $\vec{J}_{i, i+1} = (\vec{J}_i, \vec{J}_{i+1})$, $\vec{J}_i = (J_{i,x}, J_{i,Y}, J_{i,Z})$, and
\begin{equation}
    H_{j,E}(\vec{J}_j) = -J_{j,x} X_j \otimes X_E - J_{j,y} Y_j \otimes Y_E - J_{j,z} Z_j \otimes Z_E    
\end{equation}
for $j=i,i+1$, with $X_j$, $Y_j$, and $Z_j$ being the Pauli operators on the qubit $j$ ($j=i,i+1,E$). The coupling constants are randomly chosen from a Gaussian distribution.

Given this Heisenberg-type interaction between qubits $i, i+1$ and the environment qubit $E$, the noise $\cN_{p,t}$ at time $t$ for an initial state $\rho_{i, i+1,2}$ of the two qubits is given by
\begin{multline}
\cN_{i, i+1,p,t, \vec{J}}(\rho_{i, i+1}) \\
:= \tr_E \bigl[ e^{-i  t H_{i, i+1,E}(\vec{J}_{i, i+1})} (\rho_{i, i+1} \otimes \rho_E(p))  e^{i t H_{i,i+1,E}(\vec{J}_{i, i+1}) } \bigr].
\end{multline}
Thus, the whole system, which we assume to consist of $2n$ qubits, is affected by the noisy map
\begin{equation}
    \cN_{p,t, \vec{J}} = \bigotimes_{i=1}^n \cN_{2i-1, 2i,p,t, \vec{J}_{2i-1,2i}}.
\end{equation}

Based on the assumption that the dynamics in the system is information scrambling, the recovery error should satisfy \eqref{Ineq:error}. By computing the conditional max-entropy $h(p,t,\vec{J})$, which is explicitly given by
\begin{equation}
    h(p,t,\vec{J}) = \frac{1}{n}\sum_{i=1}^n h_{\cN_{2i-1, 2i,p,t, \vec{J}_{2i-1,2i}}},
\end{equation}
where $\vec{J} = ( \vec{J}_{1,2}, \dots, \vec{J}_{2n-1,2 n})$, we can estimate the recovery error for given coupling constants $\vec{J}$. We especially consider an average error over the fluctuation of the coupling constants $\vec{J}$:
\begin{multline}
\mbb{E}_{\vec{J}}\delta(c, q|\cN_{p,t,\vec{J}}, n) \\
\lesssim
\begin{cases}
\bigl( 1+ 2^{ \frac{cN}{2}} \bigr)^{1/2} \mbb{E}_{\vec{J}} 2^{(2q +  \frac{ h(p,t,\vec{J}) }{m} - 1)\frac{N}{4}}, & (c \neq 0)\\
 \mbb{E}_{\vec{J}} 2^{(2q +  \frac{ h(p,t,\vec{J}) }{m} - 1)\frac{N}{4}}, & (c=0).
\end{cases}
\end{multline}
To simplify numerical evaluations, we use the convexity of exponential functions and obtain
\begin{multline}
\mbb{E}_{\vec{J}}\delta(c, q|\cN_{p,t,\vec{J}}, n)\\
\lesssim
\begin{cases}
\bigl( 1+ 2^{ \frac{cN}{2}} \bigr)^{1/2} \mbb{E}_{\vec{J}_0} 2^{(2q +  \frac{ h_0(p,t,\vec{J}_0) }{m} - 1)\frac{N}{4}}, & (c \neq 0)\\
 \mbb{E}_{\vec{J}_0} 2^{(2q +  \frac{ h_0(p,t,\vec{J}_0) }{m} - 1)\frac{N}{4}}, & (c=0),
\end{cases} \label{Eq:last}
\end{multline}
where $h_0(p,t,\vec{J}_0)$ with $\vec{J}_0 = \vec{J}_{i_0-1,i_0}$ is the conditional max-entropy corresponding to a noisy map $\cN_{i_0-1, i_0,p,t, \vec{J}_0}$ on the qubits $i_0, i_0+1$, and an environmental qubit for a canonical index $i_0$. 

\begin{figure}[t!]
\centering
\includegraphics[width=80mm,clip]{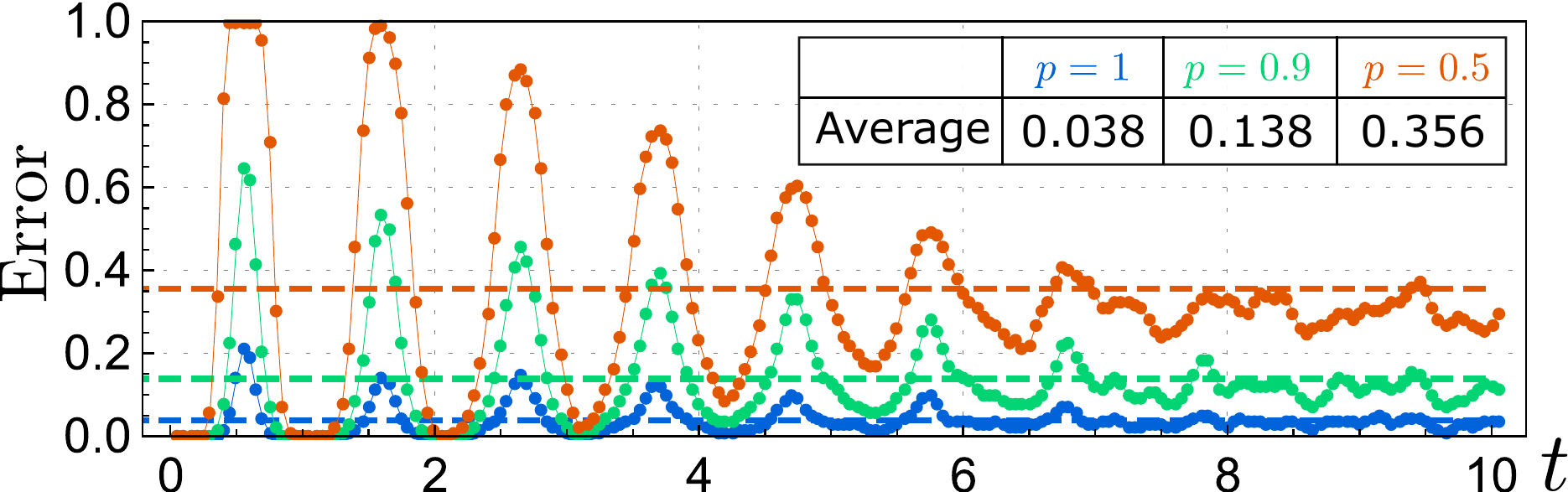}
\caption{Upper bounds on the errors on recovering quantum information in a noisy quantum chaotic system, where noise is induced from an environment at finite temperature (see the main text). The temperature is represented by $p$, i.e., $p=1$ and $0.5$ correspond to the zero and infinite temperature, respectively. The blue, green, and orange plots represent $p=1, 0.9$, and $0.5$, respectively. At each point, we have taken the average over $50$ random choices of the coupling constants $\vec{J}_0$ from a Gaussian distribution with average $1$ and standard deviation $0.25$. The dashed lines show the averages over time. Since we are interested in quantum information, we set $c=0$ and $q=0.5$. For visibility, we have chosen $n=100$. }
\label{Fig:2Qdetail}
\end{figure}

We particularly consider the situation, where the half of the particles in the system initially carries quantum information, and the rest is entangled with an auxiliary system. This corresponds to the case of $c=0$ and $q=0.5$. The initial state of the environment $E$ is set to a thermal state of the Pauli-$Z$ Hamiltonian with certain temperature.
Based on~\eqref{Eq:last}, we numerically obtain an upper bound of the average recovery error, which is depicted in Fig.~\ref{Fig:2Qdetail}.

We observe from Fig.~\ref{Fig:2Qdetail} that the average recovery error converges to a certain value depending on the initial temperature of the environment. In particular, lower temperature tends to result in small errors, which is naturally expected since the environment at higher temperature induces stronger noise into the system.
It is also worthwhile to mention that, in the thermodynamic limit, the recovery error tends to $0$ at any point strictly smaller than $1$ in the figure, and remains $1$ at the points with the error being $1$ in the figure. As clearly observed from Fig.~\ref{Fig:2Qdetail}, the recovery errors are strictly below $1$ at most of the time $t \in [0,10]$. Thus, this implies that, in a sufficiently large quantum many-body system, soon after scrambling occurs as a result of internal chaotic dynamics, the system becomes robust against any noise induced from small thermal environment.
This provides a quantitative evidence on the conjecture that QEC is intrinsic in quantum chaotic systems. 

Before we conclude, we emphasize that \eqref{Ineq:error} is applicable to any noise and also that Theorem~\ref{Thm:achiSM} can be applied to the case without assistance of entanglement. Hence, using our results, we can check the capability of QEC as a consequence of information scrambling for various situations. Information scrambling and QEC are the key to uncover the exotic relation between quantum chaos and quantum gravity. Our results provide a powerful tool to obtain more insight into scrambling, QEC, and quantum duality, in a quantitative manner.

\section{Conclusion and Discussions}  \label{S:conc}
In this paper, we have provided the one-shot hybrid capacity theorem and its applications to EAQEC by noisy quantum circuits and to quantum chaos. The theorem is represented in a highly general form, so that most of the known capacity theorems are obtained as special cases~\cite{wakakuwa2020randomized}. The theorem is provided with an explicit encoding scheme, where the CPTP maps are treated as parameters of encodings. 

In the applications, we take the advantage of the high generality of our theorem. In the first application, we have investigated EAQEC by noisy short RQCs and have revealed that a proof-of-principle demonstration by NISQ devices is possible.
In the second, we have addressed QEC in quantum chaotic systems and have provided a quantitatively support of the statement that QEC is intrinsic in quantum chaos.

For future work, it will be of significant interest to further explore the possibility of demonstrating EAQEC by NISQ devices. It is also important to generalize the noise model in our analysis of QEC in quantum chaos, which will make a conjecture about the connection between QEC and quantum chaos more solid.

\section{Acknowledgments}
YN, EW, and HY contributed equally to this work.
EW is supported by JSPS KAKENHI (Grant No.~18J01329), Japan.
YN is supported by PRESTO Grant Number JPMJPR1865, Japan.
HY is supported by  CREST (Japan Science and Technology Agency) JPMJCR1671, Cross-ministerial  Strategic Innovation  Promotion Program (SIP) (Council for Science, Technology and Innovation (CSTI)), JSPS Overseas Research Fellowships, and JST, PRESTO Grant Number JPMJPR201A, Japan.

\bibliographystyle{unsrt}
\bibliography{bibbib.bib,yamasaki_citation.bib}

\appendix

\section{Equivalence of two conditions of achievability} \label{App:EquivCond}

We show the equivalence of~\eqref{Task1} and~\eqref{Eq:fin}.
We first prove that~\eqref{Eq:fin} implies~\eqref{Task1}.
Since $\Phi_{C, Q}^{MR} = \Omega_C^{M_c R_c} \otimes \Phi_Q^{M_q R_q}$, it follows from~\eqref{Eq:fin} that
\begin{equation}
\frac{1}{2^C}\sum_j \bigl|\! \bigl| 
\tilde{\cD} \circ \cN \circ \tilde{\cE}(\ketbra{j}{j}^{M_c} \otimes \Phi_Q \otimes \Phi_E)
-
\ketbra{j}{j}^{M_c} \otimes \Phi_Q
\bigr|\! \bigr|_1 \leq \delta,
\end{equation}
where we have used the orthogonal relation in $R_c$. 
Using (\ref{eq:cEcEj}), we obtain
\begin{equation}
\frac{1}{2^C}\sum_j \bigl|\! \bigl| 
\tilde{\cD} \circ \cN \circ \tilde{\cE}_j(\Phi_Q \otimes \Phi_E)
-
\ketbra{j}{j}^{M_c} \otimes \Phi_Q
\bigr|\! \bigr|_1 \leq \delta.
\end{equation}From H\"{o}lder's inequality and the monotonicity of the trace norm under the partial trace, we have
\begin{equation}
\frac{1}{2^C}\sum_j \bigl|\! \bigl| 
\tr_{M_c} [ \ketbra{j}{j}^{M_c}\tilde{\cD} \circ \cN \circ \tilde{\cE}_j(\Phi_Q \otimes \Phi_E)]
-
\Phi_Q
\bigr|\! \bigr|_1 \leq \delta.
\end{equation}
Using (\ref{eq:cDcDj}), we arrive at
\begin{equation}
\frac{1}{2^C}\sum_j \bigl|\! \bigl| 
\tilde{\cD}_j \circ \cN \circ \tilde{\cE}_j(\Phi_Q \otimes \Phi_E)]
-
\Phi_Q
\bigr|\! \bigr|_1 \leq \delta,
\end{equation}
which is~\eqref{Task1}.

We next show the converse. 
Using (\ref{omegaCdfn}) and the orthogonal relation in $R_c$, we have
\begin{align}
&\bigl|\! \bigl| 
\tilde{\cD} \circ \cN \circ \tilde{\cE}(\Omega_C \otimes \Phi_Q \otimes \Phi_E)
-
\Omega_C \otimes \Phi_Q
\bigr|\! \bigr|_1\\
&=
\frac{1}{2^C} \sum_j
\bigl|\! \bigl| 
\tilde{\cD}_j \circ \cN \circ \tilde{\cE}_j(\Phi_Q \otimes \Phi_E)
-
\Phi_Q
\bigr|\! \bigr|_1 \notag \\
&\hspace{15mm}
+
\frac{1}{2^C} \sum_{i \neq j}
\bigl|\! \bigl| 
\tilde{\cD}_i \circ \cN \circ \tilde{\cE}_j(\Phi_Q \otimes \Phi_E)
\bigr|\! \bigr|_1\\
&\leq
\delta
+
\frac{1}{2^C} \sum_{i \neq j}
\tr[
\tilde{\cD}_i \circ \cN \circ \tilde{\cE}_j(\Phi_Q \otimes \Phi_E)
],
\end{align}
where the last line follows from~\eqref{Task1}.
The second term is calculated to be
\begin{align}
&
\frac{1}{2^C} \sum_{i \neq j}
\tr[
\tilde{\cD}_i \circ \cN \circ \tilde{\cE}_j(\Phi_Q \otimes \Phi_E)
]\\
&
=\frac{1}{2^C} \sum_{i, j}
\tr[
\tilde{\cD}_i \circ \cN \circ \tilde{\cE}_j(\Phi_Q \otimes \Phi_E)
]
\nonumber\\
&\quad\quad
-
\frac{1}{2^C} \sum_{j}
\tr[
\tilde{\cD}_j \circ \cN \circ \tilde{\cE}_j(\Phi_Q \otimes \Phi_E)
]
\nonumber\\
&=
1-
\frac{1}{2^C} \sum_{j}
\|
\tilde{\cD}_j \circ \cN \circ \tilde{\cE}_j(\Phi_Q \otimes \Phi_E)
\|_1,
\end{align}
where we have used the fact that $\{ \tilde{\cD}_j \}_j$ is the instrument. 
The triangle inequality for the trace norm and Inequality~\eqref{Task1} implies that
\begin{align}
&
\frac{1}{2^C} \sum_{j}
\|
\tilde{\cD}_j \circ \cN \circ \tilde{\cE}_j(\Phi_Q \otimes \Phi_E)
\|_1
\\
&\geq 
\frac{1}{2^C}\sum_j [\|\Phi_Q\|_1-\bigl|\! \bigl| 
\tilde{\cD}_j  \circ \cN \circ \tilde{\cE}_j(\Phi_Q \otimes \Phi_E)]
-
\Phi_Q
\bigr|\! \bigr|_1]\\
&\geq 
1-\delta.
\end{align}
Combining these all together, we arrive at (\ref{Eq:fin}).

\section{Evaluation of Errors Without Encoding}
\label{app:EEWE}

We consider the situation in which $(cN)$-bit of classical and $(qN)$-qubits of quantum information are stored into $N$ noisy qubits.
We evaluate errors when the information is directly stored in noisy qubits.
The resource of shared entanglement does not help in this case.
It is convenient to separate the problem into two cases, that is, (i) $c+q\leq 1$ and (ii) $c+q>1$.
By ``directly storing information in noisy qubits'', we mean, in Case (i), that $(c+q)N$ qubits are chosen out of the $N$ noisy qubits and the information is stored by the identity map from the source to those qubits.
For Case (ii), it means that $N$ out of the $cN$-bit and $qN$-qubit of the source are chosen and are stored to the $N$ noisy qubits by the identity map.
We assume that the storage of classical information to the noisy qubits is performed in terms of the computational basis.

We consider the dephasing noise $\mathcal{D}_{p}$ with parameter $p \in [0,1]$ and the amplitude damping noise $\mathcal{A}_{\gamma}$ with parameter $\gamma \in [0,1]$.
They are defined by 
\begin{align}
    &\mathcal{D}_{p} (\rho ) = \left(1-\frac{p}{2}\right) \rho + \frac{p}{2} Z \rho Z\\
    &\mathcal{A}_{\gamma} (\rho ) = K_0 \rho K_0^{\dagger} + K_1 \rho K_1^{\dagger},
    \label{eq:dfnAr}
\end{align}
where $Z$ is the Pauli $Z$ operator, $K_0 = |0 \rangle \langle 0| + \sqrt{1-\gamma} |1 \rangle \langle 1|$, and $K_1 = \sqrt{\gamma} |0 \rangle \langle 1|$.
The error is denoted by $\delta_0(c, q|\mathcal{N}, N)$ for the noise $\ca{N}=\mathcal{D}_{p}$ or $\mathcal{A}_{\gamma}$.
We prove that the errors are evaluated as
\begin{widetext}
\alg{
\delta_0(c,q| \cD_p, N)
&
\geq
\begin{cases}
1 -2^{-(c-1+2q)N}, & \text{when $1 \leq c \leq 2$,}\\
1- \bigl( 1- \frac{p}{2} \bigr)^{(1-c)N} 2^{-2(q+c-1)N}, & \text{when $0 \leq c< 1$ and $c+q > 1$,}\\
1- \bigl( 1 - \frac{p}{2} \bigr)^{qN}, & \text{when $0 \leq c<1$ and $c+q \leq 1$,}
\end{cases}\\
\hfill\nn
\\
\delta_0(c,q| \cA_{\gamma}, N)
&
\geq
\begin{cases}
1 - 2^{ -(c-1+ 2q)N} \bigl( 1- \frac{\gamma}{2} \bigr)^{N}, & \text{when $1 \leq c\leq 2$,}\\
1-2^{-2(q+c-1)N} \bigl(1-\frac{\gamma}{2} \bigr)^{cN} \bigl(\frac{1+\sqrt{1-\gamma}}{2} \bigr)^{2(1-c)N}, & \text{when $0 \leq c<1$ and $c+q > 1$,}\\
1 - \bigl( 1 - \frac{\gamma}{2} \bigr)^{cN}
\bigl( \frac{1+ \sqrt{1-\gamma}}{2} \bigr)^{2qN}, & \text{when $0\leq c < 1$ and $c+q \leq 1$,}
\end{cases}
}
\end{widetext}
respectively.

The proof of these inequalities will be provided in the following subsections.
For Case (ii), we adopt the strategy where classical information is preferentially stored as much as possible. 
This is because when no encoding and decoding are performed, classical information is stored with less error than quantum information. 
For the rest of the information, we can only make a guess when it is to be retrieved from the noisy qubits.
Without loss of generality, we assume the classical bit sequence and the quantum state that have not been stored but are guessed to be $i_0$ and $\varphi_0$. 
Note that the length of $i_0$ and the dimension of the Hilbert space of $\varphi_0$ depends on the values of $c$ and $q$, but it will be clear from the context.

In the proof, we will extensively use the following properties of the trace distance (see e.g.~\cite{wildetext}).
First, for any states $\rho$, $\sigma$ and $\xi$, it holds that
\alg{
\|\rho\otm\xi-\sigma\otm\xi\|_1
=
\|\rho-\sigma\|_1.
\laeq{TDPr1}
}
Second, for any states $\varrho$ and $\varsigma$ in the form of
\alg{
\varrho=\frac{1}{M}\sum_{i=1}^M\rho_i\otm\proj{i},
\quad
\varsigma=\frac{1}{M}\sum_{i=1}^M\sigma_i\otm\proj{i},
}
with $\{\ket{i}\}_{i=1}^M$ being a fixed orthonormal basis, it holds that
\alg{
\|\varrho-\varsigma\|_1
=
\frac{1}{M}\sum_{i=1}^M
\|\rho_i-\sigma_i\|_1.
\laeq{TDPr2}
}
Third, for any $0\leq\lambda\leq1$ and states $\rho,\sigma,\xi$ such that ${\rm supp}[\sigma]\subseteq{\rm supp}[\rho]\perp{\rm supp}[\xi]$, we have
\alg{
\left\|\rho-[\lambda\sigma+(1-\lambda)\xi]\right\|_1
=
\left\|\rho-\lambda\sigma\right\|_1
+(1-\lambda).
\laeq{TDPr4}
}
Forth, for any Hermitian operator $X$, it holds that
\alg{
|\!| X |\! |_1 = - \tr[X] + 2 \max_{0 \leq P \leq I} \tr[PX].
}
Hence, for any normalized pure state $\ket{\psi}$ and subnormalized positive semidefinite operator $\rho$, we have
\alg{
&
\|\proj{\psi}-\rho\|_1
\nn\\
&
\geq
-{\rm Tr}[\proj{\psi}-\rho]
+
2  \tr[\proj{\psi}(\proj{\psi}-\rho)]
\\
&
=
1+{\rm Tr}[\rho]-2\bra{\psi}\rho\ket{\psi}.
\laeq{TDPr3}
}
The maximally entangled state on a two-qubit system will be simply denoted by $\ket{\Phi}$.

\subsection{Dephasing noise}

First, we consider the case where $1 \leq c \leq 2$. In this case, the best strategy is to store the $N$-bit of classical information on the noisy $N$ qubits. 
For the rest of the information, i.e., $(c-1)N$ bits of classical and $qN$ bits of quantum information, we only make a guess when information is to be retrieved from the noisy qubits.
We hence divide $M_c$ into an $N$ qubit system $M_{c1}$ and a $(c-1)N$ qubit system $M_{c2}$, where the information in $M_{c1}$ is stored in the noisy $N$ qubits. Accordingly, $R_c$ is divided into $R_{c1} R_{c2}$.
The error is given by
\begin{equation}
\delta_0(c,q|\cD_{p}, N)
=
\frac{1}{2}
\bigl|\! \bigl|
\Omega_{cN}^{M_c R_c} \otimes \Phi_{qN}^{M_q R_q}
-
\Psi^{MR}
\bigr|\! \bigr|_1,
\end{equation}
where $\Psi^{MR}$ is a state defined by
\begin{eqnarray}
\Psi^{MR} 
= (\cD_p^{\otm N})^{M_{c1}}(\Omega_N^{M_{c1} R_{c1}}) \otimes \proj{i_0}^{M_{c2}} \otimes \pi_{(c-1)N}^{R_{c2}} 
\nn\\
\otimes \varphi_0^{M_q} \otimes \pi_{qN}^{R_q}.
\quad\quad
\end{eqnarray}
Due to the relations $(\cD_p^{\otm N})^{M_{c1}}(\Omega_N^{M_{c1} R_{c1}}) = \Omega_N^{M_{c1} R_{c1}}$, $\Omega_{cN}^{M_{c} R_{c}}=\Omega_{N}^{M_{c1} R_{c1}}\otm\Omega_{(c-1)N}^{M_{c2} R_{c2}}$ and \req{TDPr1}, we have
\alg{
2\delta_0(c,q|\cD_{p}, N)
&
=
\bigl|\! \bigl|
\Omega_{(c-1)N}^{M_{c2} R_{c2}} \otimes \Phi_{qN}^{M_q R_q} 
\nn\\
&\quad\quad
-
\proj{i_0}^{M_{c2}} \otimes \pi_{(c-1)N}^{R_{c2}} \otimes \varphi_0^{M_q} \otimes \pi_{qN}^{R_q}
\bigr|\! \bigr|_1.
\nn
}
Using
\alg{
\Omega_{(c-1)N}^{M_{c2} R_{c2}} 
&= \frac{1}{2^{(c-1)N}} \sum_{i=1}^{2^{(c-1)N}} \ketbra{i}{i}^{M_{c2}} \otimes \ketbra{i}{i}^{R_{c2}},
\\
 \pi_{(c-1)N}^{R_{c2}} 
 &= \frac{1}{2^{(c-1)N}} \sum_{i=1}^{2^{(c-1)N}} \ketbra{i}{i}^{R_{c2}},
 }
 and the relations \req{TDPr2}, \req{TDPr4} and \req{TDPr3}, it follows that
\alg{
&
2\delta_0(c,q|\cD_{p}, N)
\nn\\
&
=
\frac{1}{2^{(c-1)N}} \sum_{i=1}^{2^{(c-1)N}}
\bigl|\! \bigl|
\proj{i}^{M_{c2}} \otimes \Phi_{qN}^{M_q R_q} 
\nn\\
&\quad\quad\quad\quad\quad\quad\quad\quad
-
\proj{i_0}^{M_{c2}} \otimes \varphi_0^{M_q} \otimes \pi_{qN}^{R_q}
\bigr|\! \bigr|_1
\\
&
=
\frac{1}{2^{(c-1)N}}
\bigl|\! \bigl|
\Phi_{qN}^{M_q R_q}
-
\varphi_0^{M_q} \otimes \pi_{qN}^{R_q}
\bigr|\! \bigr|_1
+
2 \biggl( 1 -\frac{1}{2^{(c-1)N}} \biggr)
\nn
\\
&\geq
2-\frac{2}{2^{(c-1)N}}\bra{\Phi_{qN}}(\varphi_0 \otimes \pi_{qN})\ket{\Phi_{qN}}
\laeq{ann}
\\
&=
2\left(1 -\frac{1}{2^{(c-1+2q)N}}\right).
}

Second, we consider the case where $0 \leq c < 1$ and $c+q >1$. In this case, all the $cN$-bit classical information and $(1-c)N$-qubit out of $qN$-qubit quantum information are stored in the noisy $N$ qubits. 
We only make a guess for the rest of quantum information of $(c+q-1)N$ qubits.
Dividing $M_q$ into a $(1-c)N$-qubit system $M_{q1}$ and a $(c+q-1)N$-qubit system $M_{q2}$, and $R_q$ into $R_{q1}R_{q2}$, the error is given by
\begin{equation}
\delta_0(c,q|\cD_{p}, N)
=
\frac{1}{2}
\bigl|\! \bigl|
\Omega_{cN}^{M_c R_c} \otimes \Phi_{qN}^{M_q R_q}
-
\Psi^{MR}
\bigr|\! \bigr|_1,
\end{equation}
where
\begin{eqnarray}
\Psi^{MR} =( \cD_p^{\otm cN})^{M_{c}}(\Omega_{cN}^{M_{c} R_{c}}) \otimes (\cD_p^{\otm (1-c)N})^{M_{q1}}(\Phi_{(1-c)N}^{M_{q1} R_{q1}}) 
\nn\\
\otimes \varphi_0^{M_{q2}} \otimes \pi_{(c+q-1)N}^{R_{q2}}.
\quad\quad
\end{eqnarray}
Using the relations
\alg{
( \cD_p^{\otm cN})^{M_{c}}(\Omega_{cN}^{M_{c} R_{c}}) 
=\Omega_{cN}^{M_{c} R_{c}}
\laeq{rebe1}
}
and
\alg{
(\cD_p^{\otm (1-c)N})^{M_{q1}}(\Phi_{(1-c)N}^{M_{q1} R_{q1}}) 
=
([\cD_p \otm{\rm id}(\Phi)]^{\otm  (1-c)N})^{M_{q1}R_{q1}},
}
we obtain
\begin{equation}
\delta_0(c,q|\cD_{p}, N)
=
\frac{1}{2}
\bigl|\! \bigl|
 \Phi_{qN}^{M_q R_q}
-
\tilde{\Psi}^{M_qR_q}
\bigr|\! \bigr|_1,
\laeq{zetj}
\end{equation}
where
\begin{eqnarray}
\tilde{\Psi}^{M_qR_q} =
([\cD_p \otm{\rm id}(\Phi)]^{\otm  (1-c)N})^{M_{q1}R_{q1}}
\nn\\
\otimes \varphi_0^{M_{q2}} \otimes \pi_{(c+q-1)N}^{R_{q2}}.
\end{eqnarray}
Noting that
\begin{equation}
\cD_p \otimes {\rm id} (\Phi) = \bigl( 1- \frac{p}{2} \bigr) \Phi + \frac{p}{2} (Z \otimes I )\Phi (Z \otimes I ),
\end{equation}
and that $(Z \otimes I) \ket{\Phi}$ is orthogonal to $\ket{\Phi}$, the state $([\cD_p\otm{\rm id} (\Phi)]^{\otm  (1-c)N})$ is expanded as
\alg{
([\cD_p\otm{\rm id} (\Phi)]^{\otm  (1-c)N})
=
\alpha
\Phi_{(1-c)N}
+
(1-\alpha)\zeta,
\laeq{zetaj}
}
where $\alpha:=( 1- p/2 )^{(1-c)N}$ and $\zeta$ is a state that is orthogonal to $\Phi_{(1-c)N}$.
Thus, using \req{TDPr4}, \req{TDPr3}, and $\Phi_{qN}^{M_{q}R_{q}}=\Phi_{(1-c)N}^{M_{q1}R_{q1}}\otm\Phi_{(c+q-1)N}^{M_{q2}R_{q2}}$, we obtain from \req{zetj} that
\alg{
&
2\delta_0(c,q|\cD_{p}, N)
\nn\\
&
=
\left\|\Phi_{(c+q-1)N}^{M_{q2}R_{q2}}-\alpha \varphi_0^{M_{q2}} \otimes \pi_{(c+q-1)N}^{R_{q2}}\right\|_1+(1-\alpha)
\\
&\geq
2-2 \alpha \bra{\Phi_{(c+q-1)N}}(\varphi_0 \otimes \pi_{(c+q-1)N})\ket{\Phi_{(c+q-1)N}}
\!\!
\\
&=
2-
2\biggl( 1- \frac{p}{2} \biggr)^{(1-c)N} \frac{1}{2^{2(q+c-1)N}}.
}

Finally, we consider the case where $0 \leq c<1$ and $c+q \leq 1$. In this case, all the $cN$-bit and $qN$-qubit information are stored in the noisy $N$ qubits. Thus, the error is given by
\alg{
&
\delta_0(c,q|\cD_{p}, N)
\nn\\
&
=
\frac{1}{2}
\bigl|\! \bigl|
\Omega_{cN}^{M_c R_c} \otimes \Phi_{qN}^{M_q R_q}
\nn\\
&\quad\quad
-
(\cD_p^{\otm cN})^{M_c}( \Omega_{cN}^{M_c R_c}) \otimes (\cD_{p}^{\otm qN})^{M_q}(\Phi_{qN}^{M_q R_q})
\bigr|\! \bigr|_1.
\!
}
Using \req{rebe1}, this is simplified to
\begin{equation}
\delta_0(c,q|\cD_{p}, N)
=
\frac{1}{2}
\bigl|\! \bigl|
\Phi_{qN}^{M_q R_q}
-
(\cD_p^{\otm qN})^{M_q}(\Phi_{qN}^{M_q R_q})
\bigr|\! \bigr|_1.
\end{equation}
Due to the similar relation to \req{zetaj}, we arrive at
\begin{equation}
\delta_0(c,q|\cD_{p}, N)
=
1- \biggl( 1 - \frac{p}{2} \biggr)^{qN}.
\end{equation}

\subsection{Amplitude damping noise}

Let us first consider the case where $1 \leq c \leq 2$. In this case, we only store $N$-bit classical information in the noisy $N$ qubits.
Thus, we divide $M_c$ into $M_{c1}$ of $N$ qubits and $M_{c2}$ of $(c-1)N$ qubits. Accordingly, $R_c$ is divided into $R_{c1} R_{c2}$.
We define a state $\Psi^{MR}$ by
\alg{
\Psi^{MR}
\coloneqq 
(\cA_{\gamma}^{\otm N})^{M_{c1}} (\Omega_{N}^{M_{c1} R_{c1}}) \otimes \psi^{M_{c2} R_{c2} M_q R_q },
}
where
\begin{align}
\!
\psi^{M_{c2} R_{c2} M_q R_q }=\proj{i_0}^{M_{c2}} \!\otimes\! \pi^{R_{c2}}_{(c-1)N}\! \otimes\! \varphi_0^{M_q} \!\otimes\! \pi_{qN}^{R_q}.
\!
\end{align}
The error is given by
\begin{equation}
\delta_0(c,q|\cA_{\gamma}, N)
=
\frac{1}{2}
\bigl|\! \bigl|
\Omega_{cN}^{M_c R_c} \otimes \Phi_{qN}^{M_q R_q}
-
\Psi^{MR}
\bigr|\! \bigr|_1.
\end{equation}

Using $\Omega_{cN}^{M_cR_c}=\Omega_{N}^{M_{c_1}R_{c_1}} \otimes \Omega_{(c-1)N}^{M_{c_2}R_{c_2}}$ and the relation \req{TDPr2} twice, we have
\begin{widetext}
\begin{align}
&
2\delta_0(c,q|\cA_\gamma, N)
\nn\\
&=
\frac{1}{2^{N}}
\sum_{i = 0}^{2^{N}-1}
\biggl|\! \biggl|
\proj{i}^{M_{c1}} \otimes  \Omega_{(c-1)N}^{M_{c2}R_{c2}} \otimes \Phi_{qN}^{M_q R_q}
-
(\cA_{\gamma}^{\otm N})^{M_{c1}} ( \proj{i}^{M_{c1}} ) \otimes \psi^{M_{c2}R_{c2}M_qR_q} \biggr|\! \biggr|_1
\\
&=
\frac{1}{2^{N}}
\sum_{i = 0}^{2^{N}-1}
\frac{1}{2^{(c-1)N}}
\sum_{i' = 0}^{2^{(c-1)N}-1}
\biggl|\! \biggl|
\proj{i}^{M_{c1}} \!\otimes\!  \proj{i'}^{M_{c2}} \!\otimes\! \Phi_{qN}^{M_q R_q}
-
(\cA_{\gamma}^{\otm N})^{M_{c1}} ( \proj{i}^{M_{c1}} ) \otimes \proj{i_0}^{M_{c2}} \! \otimes\! \varphi_0^{M_q} \!\otimes\! \pi_{qN}^{R_q} \biggr|\! \biggr|_1
\\
&=
2 \biggl(1 - \frac{1}{2^{(c-1)N}}\biggr)
+
\frac{1}{2^{cN}}
\sum_{i = 0}^{2^{N}-1}
\biggl|\! \biggl|
\proj{i}^{M_{c1}} \otimes \Phi_{qN}^{M_q R_q}
-
(\cA_{\gamma}^{\otm N})^{M_{c1}}  ( \proj{i}^{M_{c1}} ) \otimes \varphi_0^{M_q} \otimes \pi_{qN}^{R_q}
\biggr|\! \biggr|_1. \label{Eq:rrre}
\end{align}
\end{widetext}
The equality (\ref{Eq:rrre}) is directly obtained from \req{TDPr1} and the fact that both $\proj{i_0}^{M_{c2}} \otimes \pi_{(c-1)N}^{R_{c2}}$ and $\Omega_{(c-1)N}^{M_{c2} R_{c2}}$ are diagonal in the computational basis.
To evaluate the trace distance in (\ref{Eq:rrre}), note that
 for any bit-sequence $i$ with length $N$, it holds that 
\alg{
\cA_{\gamma}^{\otimes N} (\proj{i}) = (1-\gamma)^{N - z_i} \proj{i} + \bigl[1- (1-\gamma)^{N - z_i} \bigr] \rho_{\neq i}.
\laeq{rccx12}
}
Here, $z_i$ is the number of zero in $i$, and $\rho_{\neq i}$ is a normalized state that does not have a support on $\ketbra{i}{i}$. 
Thus, using \req{TDPr4}, each term in the summation in~\eqref{Eq:rrre} is calculated to be 
\begin{equation}
\bigl|\! \bigl|
\Phi_{qN}^{M_q R_q}
-
(1-\gamma)^{N - z_i}  \varphi_0^{M_q} \otimes \pi_{qN}^{R_q}
\bigr|\! \bigr|_1
+
1- (1-\gamma)^{N - z_i}. \label{Eq:rrreee}
\end{equation}

The first term in~\eqref{Eq:rrreee} is bounded from below by using the relation \req{TDPr3}, leading to
\begin{multline}
\bigl|\! \bigl|
\Phi_{qN}^{M_q R_q}
-
(1-\gamma)^{N - z_i}  \varphi_0^{M_q}\otimes \pi_{qN}^{R_q}
\bigr|\! \bigr|_1\\
\geq 
1 + (1-\gamma)^{N - z_i} - 2 (1-\gamma)^{N - z_i}2^{- 2qN}.
\end{multline}
Substituting these into~\eqref{Eq:rrre}, we arrive at
\begin{align}
&
\delta_0(c,q|\cA_{\gamma}, N)
\nn\\
&\geq 
1 - \frac{1}{2^{(c-1)N}}
+
\frac{1}{2^{cN}}
\sum_{i = 0}^{2^{N}-1}
[1 -  (1-\gamma)^{N - z_i}2^{- 2qN}]
\\
&=
1 - \frac{1}{2^{ (c-1 + 2q)N}} \biggl( 1- \frac{\gamma}{2} \biggr)^{N},
\end{align}
where the last line follows from the relation that
\alg{
\sum_{i=0}^{2^N-1} (1-\gamma)^{N - z_i} = (2-\gamma)^N.
\laeq{rebe}
}

We next consider the case 
 where $0 \leq c < 1$ and $c+q \geq 1$.
 In this case, all the classical information and $(1-c)N$-qubit quantum information are stored in the noisy $N$ qubits. We hence divide $M_q$ and $R_q$ into $M_{q1}M_{q2}$ and $R_{q1}R_{q2}$, respectively. The $M_{q1}$ consists of $(1-c)N$ qubits to be stored. We define the state $\Psi$ by
\alg{
\Psi^{MR}= (\cA_{\gamma}^{\otm cN})^{M_c}(\Omega_{cN}^{M_c R_c}) \otimes \psi^{M_qR_q},
}
where
\begin{equation}
\psi^{M_qR_q}
=
(\cA_{\gamma}^{\otm (1-c)N})^{M_{q1}}(\Phi_{(1-c)N}^{M_{q1} R_{q1}})  \otimes \varphi_0^{M_{q2}} \otimes \pi_{(q+c-1)N}^{R_{q2}}.
\laeq{dfnpsipsi}
\end{equation}
The error is represented as
\begin{equation}
\delta_0(c,q|\cA_{\gamma}, N)\\
=
\frac{1}{2}
\bigl| \! \bigl|
\Omega_{cN}^{M_cR_c} \otimes \Phi_{qN}^{M_qR_q}
-
\Psi^{MR}
\bigr| \! \bigr|_1.
\end{equation}
From \req{TDPr2}, it follows that
\alg{
&
2\delta_0(c,q|\cA_{\gamma}, N)
\nn\\
&
=
\frac{1}{2^{cN}} \sum_{i=0}^{2^{cN}-1}
\bigl| \! \bigl|
\proj{i}^{M_c} \otimes \Phi_{qN}^{M_qR_q}
\nn\\
&\quad\quad\quad\quad\quad\quad
-
(\cA_{\gamma}^{\otm cN})^{M_c}(\proj{i}^{M_c}) \otimes \psi^{M_qR_q}
\bigr| \! \bigr|_1.
}

Using \req{TDPr4} and a relation similar to \req{rccx12}, each term in the summation is evaluated to be
\begin{align}
&\bigl| \! \bigl|
\proj{i}^{M_c} \otimes \Phi_{qN}^{M_qR_q}
-
(\cA_{\gamma}^{\otm cN})^{M_c}(\proj{i}^{M_c}) \otimes \psi^{M_qR_q}
\bigr| \! \bigr|_1
\nn\\
&=
\bigl| \! \bigl|
\Phi_{qN}^{M_qR_q}
-
(1- \gamma)^{cN - z_i} \psi^{M_qR_q}
\bigr| \! \bigr|_1
+
1-(1- \gamma)^{cN - z_i}\\
&\geq
2
\bigl(
1-
(1- \gamma)^{cN - z_i} \bra{\Phi_{qN}} \psi \ket{\Phi_{qN}}
\bigr),
\laeq{urayasu}
\end{align}
where we have used \req{TDPr3} in the last line.
Using \req{dfnpsipsi}, the second term in \req{urayasu} is further calculated to be
\alg{
&
\bra{\Phi_{qN}} \psi \ket{\Phi_{qN}}
\nn\\
&
=
\bra{\Phi_{(1-c)N}}
(\cA_{\gamma}^{\otm (1-c)N})^{M_{q1}}(\Phi_{(1-c)N}^{M_{q1} R_{q1}}) \ket{\Phi_{(1-c)N}}
 \cdot 
 \nn\\
 &\quad\quad
 \bra{\Phi_{(q+c-1)N}}(\varphi_0^{M_{q2}} \otimes \pi_{(q+c-1)N}^{R_{q2}}) \ket{\Phi_{(q+c-1)N}}
 \nn\\
&
=
2^{-2(q+c-1)N}
 \cdot 
 \nn\\
 &\quad\quad
\bra{\Phi_{(1-c)N}}
(\cA_{\gamma}^{\otm (1-c)N})^{M_{q1}}(\Phi_{(1-c)N}^{M_{q1} R_{q1}}) \ket{\Phi_{(1-c)N}}.
}
From the definition of $\ca{A}_\gamma$ given by \req{dfnAr}, we have
\begin{align}
&
\bra{\Phi_{(1-c)N}} (\cA_{\gamma}^{\otm (1-c)N})^{M_{q1}} (\Phi_{(1-c)N}^{M_{q1} R_{q1}}) \ket{\Phi_{(1-c)N}}
\nn\\
&=
(\bra{\Phi} (\cA_{\gamma}\otm{\rm id}) (\proj{\Phi}) \ket{\Phi})^{(1-c)N}\\
&=
\left(
\sum_{l=0,1}
|\bra{\Phi}(K_l\otm I)\ket{\Phi}|^2
\right)^{(1-c)N}
\\
&=
\left(
\sum_{l=0,1}
\left(
\frac{1}{2}
|{\rm Tr}[K_l]|
\right)^2
\right)^{(1-c)N}
\\
&=
\biggl( \frac{1+ \sqrt{1-\gamma}}{2} \biggr)^{2(1-c)N}.
\laeq{rrr1}
\end{align}
Combining these all together, and by using \req{rebe}, we obtain
\begin{multline}
\delta_0(c,q|\cA_{\gamma}, N)\\
\geq
1
-
\biggl(1-\frac{\gamma}{2} \biggr)^{cN}
\biggl(\frac{1+\sqrt{1-\gamma}}{2} \biggr)^{2(1-c)N}
\frac{1}{2^{2(q+c-1)N}}.
\nn
\end{multline}

In the last case where $0 \leq c< 1$ and $c+q \leq 1$, all information, both classical and quantum, are stored into the noisy $N$ qubits.
Hence, the error is given by
\alg{
&
\delta_0(c,q|\cA_{\gamma}, N)
\nn\\
&
=
\frac{1}{2}
\bigl|\! \bigl|
\Omega_{cN}^{M_c R_c} \otimes \Phi_{qN}^{M_q R_q} 
\nn\\
&\quad\quad\quad\;
-
(\cA_{\gamma}^{\otm cN})^{M_c} (\Omega_{cN}^{M_c R_c}) \otimes (\cA_{\gamma}^{\otm qN})^{M_q} (\Phi_{qN}^{M_q R_q})
\bigr|\! \bigr|_1.
\nn
}
Using \req{TDPr2}, we have
\alg{
&
2\delta_0(c,q|\cA_{\gamma}, N)
\nn\\
&
=
\frac{1}{2^{cN}}
\sum_{i=0}^{2^{cN}-1}
\bigl|\! \bigl|
\proj{i}^{M_c} \otimes \Phi_{qN}^{M_q R_q}
\nn\\
&\quad\quad\;\quad\quad\;
-
(\cA_{\gamma}^{\otm cN})^{M_c} (\proj{i}^{M_c}) \otimes (\cA_{\gamma}^{\otm qN})^{M_q} (\Phi_{qN}^{M_q R_q})
\bigr|\! \bigr|_1.
\nn
}

The classical part of the error is evaluated in a similar way to (\ref{Eq:rrreee}), which results in
\begin{multline}
2\delta_0(c,q|\cA_{\gamma}, N)\\
=
\frac{1}{2^{cN}}
\sum_{i=0}^{2^{cN}-1}
\biggl[
\bigl|\! \bigl|
\Phi_{qN}^{M_q R_q}
-
(1-\gamma)^{cN - z_i} (\cA_{\gamma}^{\otm qN})^{M_q} (\Phi_{qN}^{M_q R_q})
\bigr|\! \bigr|_1\\
+
1 - (1-\gamma)^{cN - z_i}  \biggr].
\end{multline}
Here, $z_i$ is the number of zeros in the $cN$-bit sequence $i$.
Using \req{TDPr3} and \req{rebe}, we obtain
\alg{
&
\delta_0(c,q|\cA_{\gamma}, N)
\nn\\
&
\geq
1 - \biggl( 1 - \frac{\gamma}{2} \biggr)^{cN}
\bra{\Phi_{qN}} (\cA_{\gamma}^{\otm qN})^{M_q} (\Phi_{qN}^{M_q R_q}) \ket{\Phi_{qN}}.
}
In the same way as \req{rrr1}, the second term is calculated to be
\begin{align}
\bra{\Phi_{qN}} (\cA_{\gamma}^{\otm qN})^{M_q} (\Phi_{qN}^{M_q R_q}) \ket{\Phi_{qN}}
=
\biggl( \frac{1+ \sqrt{1-\gamma}}{2} \biggr)^{2qN}.
\end{align}
We thus arrive at
\begin{equation}
\delta_0(c,q|\cA_{\gamma}, N)
\geq
1 - \biggl( 1 - \frac{\gamma}{2} \biggr)^{cN}
\biggl( \frac{1+ \sqrt{1-\gamma}}{2} \biggr)^{2qN}.
\end{equation}

\end{document}